\documentclass[traditabstract]{aa} 
\pdfoutput=1

\usepackage{graphicx}
\usepackage{amsmath}
\usepackage{txfonts}
\usepackage{natbib}
\usepackage{multirow}
\usepackage{longtable}
\usepackage{lscape}
\usepackage{hyperref}
\usepackage{array}
\newcolumntype{L}[1]{>{\raggedright\let\newline\\\arraybackslash\hspace{0pt}}m{#1}}
\newcolumntype{C}[1]{>{\centering\let\newline\\\arraybackslash\hspace{0pt}}m{#1}}
\newcolumntype{R}[1]{>{\raggedleft\let\newline\\\arraybackslash\hspace{0pt}}m{#1}}

\def\Ha{{H$\alpha$}}

\def\niiHa{[\ion{N}{ii}]/{H$\alpha$}}
\def\oii{[\ion{O}{ii}]}
\def\niiHa{[\ion{N}{ii}]/{H$\alpha$}}

\begin{document}

   \title{Multi--wavelength landscape of the young galaxy cluster RXJ1257.2+4738 at z=0.866}
   \subtitle{II. Morphological properties}

   \author{I. Pintos-Castro  \inst{1,2,3,4}
          \and
         M. Povi\'c \inst{5}
          \and
         M. S\'anchez-Portal \inst{6,4}
          \and
         J. Cepa \inst{3,2}
          \and
         B. Altieri \inst{6}
          \and
         \'A. Bongiovanni \inst{2,3}
          \and
         P.A. Duc \inst{7}
          \and
         A. Ederoclite \inst{8} 
          \and
         I. Oteo \inst{9,10}
          \and
         A.M. P\'erez Garc\'ia \inst{2,3}
          \and
         R. P\'erez Mart\'inez \inst{6,4}
          \and
         J. Polednikova \inst{2,3}
          \and
         M. Ram\'on--P\'erez \inst{2,3}
          \and
         S. Temporin \inst{11}
}

   \institute{Centro de Astrobiolog\'ia, INTA-CSIC, Villanueva de la Ca\~nada, Madrid, Spain  \\
         \email{ipintos@cab.inta-csic.es}
         \and
             Instituto de Astrof\'isica de Canarias, La Laguna, Tenerife, Spain
         \and
             Departamento de Astrof\'isica, Facultad de F\'isica, Universidad de La Laguna, La Laguna, Tenerife, Spain
	     \and
	        ISDEFE, Madrid, Spain.
         \and
            Instituto de Astrof\'isica de Andaluc\'ia (IAA-CSIC), Granada, Spain
         \and
         	European Space Astronomy Centre (ESAC), Villanueva de la Ca\~nada, Madrid, Spain
         \and
            Laboratoire AIM Saclay, CEA/IRFU, CNRS/INSU, Universit\'e Paris Diderot, France
         \and
	        Centro de Estudios de F\'isica del Cosmos de Arag\'on, Teruel, Spain
         \and
            Institute for Astronomy, University of Edinburgh, Royal Observatory, Edinburgh, UK
         \and
            European Southern Observatory, Garching, Germany
         \and
            Institute of Astro- and Particle Physics, University of Innsbruck, Austria
            }

  \date{Received month day, year; accepted month day, year}

  \abstract
{The study of the evolution of the morphological distribution of galaxies in different environments can provide important information about the effects of the environment and the physical mechanisms responsible for the morphological transformations. As part of a complete analysis of the young cluster RXJ1257+4738 at z\,$\sim$\,0.9, we studied in this work the morphological properties of its galaxies. We used non--parametric methods of morphological classification, as implemented in the galSVM code. The classification with the applied method was possible even using ground--based observations, and $r'$--band imaging from OSIRIS/GTC. We defined very conservative probability limits, taking into account the probability errors, in order to obtain a trustworthy classification. In this way we were able to classify $\sim$\,30\% of all cluster members, and to separate between late--type (LT) and early--type (ET) galaxies. Additionally, when analysing the colour--magnitude diagram, we observed a significant population of blue ET galaxies between the classified ones. We discussed possible explanations for the finding of this population. Moreover, we studied different physical properties of LT, ET, and blue ET galaxies. They turn out to be comparable, with the exception of the stellar mass that shows that the red ET population is more massive. We also analysed the morphology--density and morphology--radius relations, observing that, only when considering the morphological separation between ET and LT galaxies, a mild classical behaviour is obtained. RXJ1257+4738 is a young galaxy cluster, showing a clumpy structure and being still in the process of formation, which could explain the lack of some of the standard morphological relations. This  makes this cluster a very attractive case for obtaining the higher resolution data and for studying in more details the morphological properties of the entire cluster and relation with the environment. A full catalogue of 1/3 cluster galaxies with reliable morphological classification is available in the electronic version of this paper. }

               
    \authorrunning{I. Pintos-Castro et al.}          
	\titlerunning{Morphology of RXJ1257.2+4738}
	\maketitle

\section{Introduction}
\label{sec:intro}

The cores of nearby clusters are dominated by red, massive and passive elliptical and S0 galaxies whereas there are significant changes in galaxy properties towards the outer parts of clusters, notably the increase in the percentage of late-type galaxies. This observational fact was early identified \citep[e.g. ][]{Zwicky1942}. \cite{Dressler1980} found an increase of the fraction of early-type (ET; elliptical and S0 galaxies) with the local galaxy density and quantified this as the morphology-density relation (MDR). Likewise, a decrease of the fraction of star forming galaxies is observed with local density \citep[e.g. ][]{Yo2013,Webb2013,Koyama2011}. These relations evolve with the cosmic time: a significant increase in the amount of blue cluster galaxies is observed at z\,$>$\,0.2 \citep[the Butcher-Oemler effect ][]{Butcher1984}, as does the cluster star formation (SF) and AGN activity \citep[e.g.][]{Haines2009,Martini2013}. \cite{Elbaz2007} studied the SF--density relation using deep \textit{Spitzer} observations of the GOODS field, witnessing a vigorous SF activity in the centres of groups at z\,$\sim$\,1, suggesting a reversal of the SF--density relation. \cite{Tran2010} found a dramatic increase of the fraction of SF galaxies in a \textit{Spitzer}--selected cluster at z\,$\sim$\,1.62, CIG J0218.3-0501, growing by a factor of three from the lowest to the highest galaxy density regions. These evidences have been also supported by simulations from \cite{Tonnesen2014}. However, there is some controversy since \cite{Ziparo2014} found no clear evidence of such a reversal when studying the evolution of the SF--density relation in the ECDFS and GOODS fields up to z\,$\sim$\,1.6.

The correlation of the morphology and/or SF activity of the clusters' galaxies with the local density (as traced by the local galaxy density or dark matter density) can provide clues on the stage of infall at which galaxies experience the bulk of their transformations. Hence, it is important to perform wide-area studies in clusters spanning a range of redshifts.

\citet{Holden2007} performed a morphological study of 674 spectroscopically confirmed cluster galaxies from five massive clusters in the redshift range z\,=\,0.023 to 0.83 using \textit{Hubble Space Telescope} (HST) imaging. They performed the analysis using both a luminosity-limited sample (up to M$_V$\,$<$\,M$^*_V$\,$+$\,1) and a mass-selected sample (down to 10$^{10.6}$\,M$_{\odot}$) in order to avoid biases due to changes in the $M/L$ relation related to changes in star formation rate (SFR). For the luminosity-selected sample, the authors found an increasing fraction of ET galaxies with the local density. This is observed in all studied clusters, although there is a clear trend of decreasing the fraction of ET galaxies with redshift, in agreement with previous studies \citep[e.g.][]{Postman2005}. When studying the mass-selected sample, they found that the MDR still exists but shows a lower slope. However, no substantial variation of the slope with redshift was found, consistent with no evolution over a period of 7\,Gyr. The authors concluded that the redshift evolution of the MDR observed in luminosity-selected samples should be driven by galaxies with stellar masses below 10$^{10.6}$\,M$_{\odot}$.

\citet{Nantais2013} performed a luminosity-limited morphological study of 124 spectroscopically confirmed members of the cluster RX J1052.7-1357 at redshift 0.84. using HST $riz$ ACS imaging. The authors derived the morphology of the galaxies using a parametric method (S\'ersic index) and separated their sample into ET and late-type (LT; spirals) categories and a third class encompassing peculiar (e.g. galaxies with a nucleus bluer than the outer regions), compact and merging objects (hereafter referred as PEC).  They found that the global percentage of ET galaxies is 47\%, about 2.8 times higher than the field. They traced the dependency of morphology on the local dark-matter density (DMD), defining three DMD ranges: low (LDMD), intermediate (IDMD) and high (HDMD). 
In the LDMD environment, the fraction of ET galaxies is 27\% (approximately 1.5 times the value of z\,$\sim$\,0.9 field and similar to groups at the same redshift), the percentage of LT galaxies 19\% (lower than field, 31\%) and finally that of PEC galaxies is 53\%, similar to field values. In the IDMD environment, the percentage of ET galaxies raises to 61\%, that of LT galaxies decreases only very slightly to 16\% but the fraction of PEC objects goes down by half, to 23\%. Finally, in the JDMD medium the amount of ET galaxies increases to 71\%, no LT galaxies are found but the fraction of PEC galaxies is roughly maintained, 21\%. The authors suggested that in the IDMD environment, the preferred mechanism is the direct transformation of PEC galaxies into ET ones rather than from PEC to LT and then to ET. In the transition from IDMD to HDMD environment, LT galaxies disappear, transforming into ET ones. Nevertheless, this last result is less robust due to the low-number statistics and the fact that, depending on the orbit  of the galaxy, the region where the bulk of transformations take place is not necessarily that in which the galaxy is observed.

The vast majority of studies of the morphology of cluster galaxies are  carried out with HST imaging, specially at intermediate and high redshift, due to the high imaging resolution and depth required. Parametric and/or visual classification methods are customarily used. In this paper, we attempt to perform a morphological study of an intermediate redshift cluster using images from a ground--based telescope. \citet{HuertasCompany2009b} performed a morphological study of galaxies in a sample of nine rich intermediate redshift clusters in the range 0.4\,$<$\,z\,$<$\,0.6 observed with the CFHT MegaCam in the $griz$ bands up to a cluster-centric distance of some 5\,Mpc. Selecting the cluster candidates by photometric redshifts, the authors determined a broad morphological class (early/late) by means of the galSVM code \citep[][see a short description in Sect. \ref{sec:morphoClas_method}]{HuertasCompany2008} for nearly 4000 objects. The agreement between this ground-based classification and that derived from HST data is better than 90\%. When considering the whole sample, they found a a clear correlation between morphology and local density with early--type galaxies dominating the densest regions. This correlation holds not only in the central region but also in the cluster outskirts, suggesting that the morphological evolution is mainly driven by the local environment through galaxy--galaxy interactions, independent of cluster properties. Here we will apply the same approach to our data. As part of the GLACE project \citep{SanchezPortal2015}, an OSIRIS/GTC program to study the evolution of emission--line galaxies in clusters, we are performing a complete study of the young massive galaxy cluster RXJ1257.2+4738 (hereafter referred to as RXJ1257) at z\,=\,0.866. The first part of the work, reported in \cite{Yo2013} (hereafter Paper I), included the compilation of a multi--wavelength dataset to build a reliable sample of cluster members, and the study of the population of far--infrared (FIR) emitters, in order to analyse the relation of the environment with the star--formation activity and other galaxy properties as the stellar mass. In the second part, described in this paper, we have exploited the OSIRIS/GTC broad--band imaging to perform, using non--parametric methods, a morphological classification of the full cluster sample and investigate the relation between the morphology of the galaxies and their cluster environment. The third part, addressed in Pintos-Castro et al. (in preparation, hereafter Paper III), will describe the \oii\ TF/OSIRIS observations, including a catalogue of fainter star--forming galaxies formed by the RXJ1257 \oii--emitters, with the aim of building a deeper picture of the star--formation activity in the cluster, and taking the opportunity to compare both \oii\ and FIR star--formation estimators.
    
    This paper is organised as follows: in Sect.\,\ref{sec:data} we give a summary of the observations used in this study, in Sect.\,\ref{sec:morphoClas_method} we present a detailed description of the morphological classification method employed: the galSVM code, and in Sect.\,\ref{sec:morphoClas_results} we present the results obtained with this classification technique using ground--based observations. The physical properties of the different morphological types are described in Sect.\,\ref{sec:properties}, and the specific discussion about the explanations for the finding of a large optically blue early--type population are summarised in Sect.\,\ref{sec:blueET}. The investigation of the morphology--density relation in the RXJ1257 cluster is discussed in Sect.\,\ref{sec:tsigmarelation}, and a summary of the essential results is provided in Sect.\,\ref{sec:conclusions}. Throughout this paper we assume an Universe with H$_0$\,=\,70\,km\,s$^{-1}$\,Mpc$^{-1}$, $\Omega_{\Lambda,0}$\,=\,0.7 and $\Omega_m,0$\,=\,0.3. Magnitudes are listed in the AB system \citep{Oke1983}.

	\section{Data}
    \label{sec:data}
    Most of the data used in this paper were extensively described in Paper I, with the exception of the \oii\ data that will be fully detailed in Paper III. In this section we summarize the data which concern to the morphological analysis.
   
    \subsection{Optical broad--band imaging}
\begin{figure}
\centering
\includegraphics[width=\hsize]{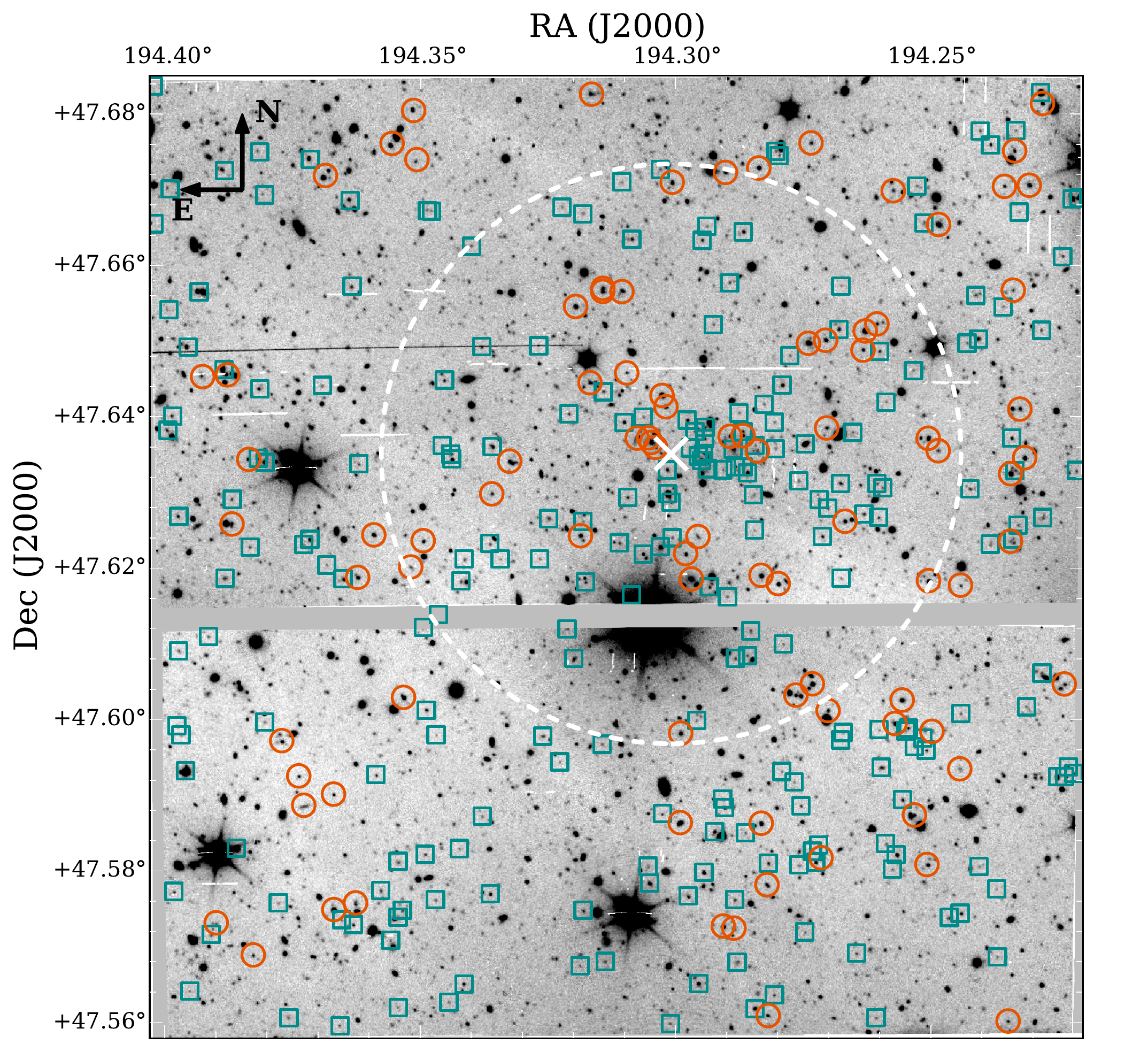}
\caption{Projected spatial distribution for the cluster galaxies, the morphological classified (orange circles) and unclassified (cyan squares) sub--samples, represented over the $r'$--band image. The white cross and dashed white circle are the cluster centre and virial radius, respectively.}
\label{fig:rband}
\end{figure}

    Broad--band imaging with SDSS $g'r'i'z'$ filters was obtained using the OSIRIS \citep{Cepa2003} instrument at the 10.4\,m GTC telescope, as part of a large ESO/GTC program (ID.\,186.A-2012, P.I.\,M.\,S\'anchez-Portal). The data were gathered during two dark nights under photometric conditions, with a seeing below 1.1$''$. Observations were planed to place the cluster centre at the CCD2, placing a bright star in the gap between both CCDs (see Fig.\,\ref{fig:rband}). Therefore, the three position dithering pattern was designed in order not to cover the gap. Total on--source exposure times for $g'r'i'z'$ were 180\,s, 450\,s, 450\,s, and 2000\,s, respectively. The reduction process was performed following standard IRAF\footnote{IRAF is distributed by the National Optical Astronomy Observatory, which is operated by the Association of Universities for the Research in Astronomy, Inc., under cooperative agreement with the National Science Foundation (\url{http://iraf.noao.edu/)}} procedures. Individual frames were bias subtracted, flat--fielded, fringing corrected (only for $z'$), aligned, and coadded to produce deep mosaics. Astrometry was performed using the SDSS DR6, obtaining an accuracy better than 0.17$''$. The flux calibration was made by comparing observations of photometric standard stars with the SDSS DR6. 
    
    Source detection and photometry were carried out using SExtractor \citep{Bertin1996}. The total flux for each source was estimated with the FLUX\_AUTO parameter, while the FLUX\_APER was used for calculating optical colours. Since we will use the $r'$--band to estimate the morphological parameters, we used this image as reference for object detection in the SExtractor dual mode. The final optical catalogue includes 1894 sources with completeness levels at 50\% of 25.0, 24.3, 24.2, and 23.4\,mag for $g'r'i'z'$, respectively. A final check of our photometry was performed against that from  \cite{Ulmer2009}. To that end, we compared the distributions of galaxies in the $i'$\,-\,$z'$ vs. $i'$ colour--magnitude diagram (a direct comparison of individual galaxies was not possible since the referred authors did not publish a source catalogue). Both distributions were found highly compatible in both magnitude and colour range and also in the relative source counts in each colour and magnitude bin.

    \subsection{OSIRIS/TF imaging: \oii}
    OSIRIS \oii\,$\lambda$3727 observations were carried out as part of the same large ESO/GTC program, during semesters 11A, 12A, and 13A, under photometric and no photometric nights. We designed the observations to cover the spectral range 6884\,-\,7114\,\AA\ (to scan a velocity field of at least $\pm$3000\,km\,s$^{-1}$) with 24 scan steps spaced 10\,\AA, with a total on--source exposure time of 18\,hours. At each TF tuning, three individual exposures of 900\,s without dithering pattern were taken (in order to avoid wavelength shifts and saturation problems). To carry out the TF data flux calibration we obtained long--slit spectroscopy of two stars in the RXJ1257 cluster field with the OSIRIS instrument. These observations were carried out under photometric conditions. For each star one spectrum of 1500\,s was taken using the R1000R grism with a slit of 2.5$''$ width and 8.67$'$ length in the spatial direction. For more details on the reduction and calibration processes of the \oii\ TF data see \cite{SanchezPortal2015} and Paper III.
    
    After the data reduction, we have 4478 pseudo--spectra available in the RXJ1257 field to select the \oii\ cluster emitters. The process for selecting the \oii\ cluster members includes: (\textit{i}) automatic classification by the definition of the existence of an emission--line in the pseudo--spectra based on a signal--to--noise criterion (1080 emission--line galaxy (ELG) candidates); (\textit{ii}) visual inspection of the pseudo--spectra automatically classified (271 ELG candidates); (\textit{iii}) visual inspection of the galaxy stamps in each TF image (161 ELGs); (\textit{iv}) rejection of interlopers using the colour--colour diagram (118 \oii--emitters); and, (\textit{v}) constraint of the line--of--sight velocity ($v_{LoS}$) covered over the entire OSIRIS field--of--view (87 
    \oii--emitters within a complete v$_{LoS}$). 
    
    \subsection{Multi--wavelength catalogue}
    In order to define a reliable sample of cluster members, the first step is to build a multi--wavelength optical--to--NIR catalogue. As already explained in Paper I, we cross--matched the optical catalogue with J--band and IRAC four channels catalogues, using the nearest--neighbour technique following \cite{Geach2006} and the \cite{Ruiter1977} methodology. This catalogue was used to estimate photometric redshifts by SED--fitting the stellar part of the spectrum with \cite{Bruzual2003} templates using the \textit{LePhare} code \citep{Ilbert2006}. The galaxies cluster membership was determined through Monte Carlo simulations: (i) we created 600 mock catalogues by randomly varying every band flux within its error; (ii) we fitted the SED with \textit{LePhare}, obtaining both a photometric redshift distribution (with 600 values) for each source and 600 photometric redshift distributions for the cluster; (iii) with the cluster distribution we defined an initial cluster redshift range, then we considered as cluster candidates those sources satisfying that 1$\sigma$ width centred at the central position of the best--fitted Gaussian function to the z$_{PHOT}$ distribution is completely included within this range; and, (iv) finally the sample was constrained redefining the cluster photometric redshift range with the accuracy ($\sigma_{\Delta z/1+z}$\,=\,0.094 at the cluster redshift), obtaining 271 galaxies with photometric redshift between 0.79 and 0.98. For more details on the SED--fitting and cluster sample definition see Paper I.
    
    The photometric cluster sample thus defined, along with a small spectroscopic sample \citep{Ulmer2009}, yielded to a cluster sample of 271\,+\,21\,=\,292 galaxies. We merged this first cluster catalogue, using the nearest neighbour technique with a maximum error radii of 1.5$''$, with the \oii--emitters sample v$_{LoS}$ complete and obtained a final cluster catalogue of 315 members. The FIR data, already exploited in the first paper, was again matched, just as explained in \citet{Yo2013}, with this updated cluster catalogue. Therefore, the final catalogue of 315 cluster galaxies used in this work includes: 21 spec--z, 87 TF--z with \oii\ emission, and 207 photo--z. This sample also includes 37 FIR emitters.


\section{Morphological classification: methodology}
\label{sec:morphoClas_method}

Accomplishing the morphological classification of galaxies in the RXJ1257 cluster is a complex task. First, because of its high redshift distribution, and second, since, up to now, high-resolution optical data are unavailable. Therefore, the visual classification and methods based on galaxy decomposition and fitting were discarded from the beginning, and we instead decided to test the non-parametric methods. 

In this work we used the galSVM code\footnote{http://gepicom04.obspm.fr/galSVM/Home.html}  \citep{HuertasCompany2008}. It is a publicly available code that works in IDL, and uses the freely available library libSVM \citep{Chang2001}, to perform morphological classification in an automated way using support vector machines (SVM)\footnote{Group of supervised learning methods that can be applied to classification or regression.}. This code can use a number of parameters simultaneously and non-linear boundaries between them to classify galaxies. Moreover, it was already tested on high--redshift and low-resolution samples of galaxies \citep[e.g.][and references there in]{HuertasCompany2008, HuertasCompany2009a, HuertasCompany2011, HuertasCompany2014temp, Povic2013}, obtaining reliable classifications. 

We followed the standard galSVM procedure, as described in \citet{HuertasCompany2008}. The code uses a training set of local galaxies with known visual morphology to train the SVM that is then applied to classify the real dataset (RXJ1257 cluster galaxies in our case). galSVM creates simulated galaxies from the local sample ones: galaxies that are redshifted and scaled in luminosity to match the magnitude and redshift distributions of the studied sample, re-sampled with the RXJ1257 $r'$ broad--band pixel scale, and dropped in a real cluster image background. A set of morphological parameters is then measured on the local dataset of simulated galaxies, and after on the cluster sample, and used to train the vector machine. galSVM code analyses the simulated sample by comparing the obtained morphological parameters and the original visual classification, and performs a weighted distribution of the morphological type for the real cluster sample. During this last step, the support vector--based learning machine is trained with a fraction of the simulated sample, so that when repeating the process through Monte--Carlo simulations the sub--sample of the simulated sample is different in each run. From each Monte--Carlo simulation we obtain one probability, with values  between 0 and 1, for every galaxy to be ET ($p_{ET}$). The final probability was obtained as average of all previous probabilities. The probability for the source to be LT is then $p_{LT}$\,=\,1\,-\,$p_{ET}$. We assumed that there is no significant change in galaxy properties between the local and high--redshift samples, although brightness dimming is taken into account. This assumption might be strong for our redshift $\sim$\,0.9, but to minimize the effect we classified galaxies into two broad morphological types (ET and LT), and for each Monte-Carlo run we forced galSVM to select the same number of ET and LT local galaxies from the training sample. Finally, in our two--type classification the ETs include elliptical (E), lenticular (S0), and S0a galaxies, while the LT group consists of spirals and irregulars.

	\subsection{Description of the measured morphological parameters}
    \label{morph_parameters}
    
We measured six morphological parameters, commonly used in the non--parametric methods. In the following we provide a brief  description for each of them. In all definitions, when necessary, the galaxy centre is determined by minimizing the asymmetry index, while the total flux is defined as the one contained within 1.5 times the Petrosian radius \citep[$r_p$, measured by SExtractor; see][]{HuertasCompany2008}.

\begin{itemize}

\item[$\clubsuit$] \textbf{Abraham concentration index}  \citep[C,][]{Abraham1996}. It is defined as the ratio between the fluxes of two isophotes. In our case the inner and outer isophotes contain 30\% and 90\% of the total flux, respectively:

\begin{equation}
C\,=\,\frac{\sum\,\sum\limits_{i,j \in E(0.3)}\,I(i,j)}{\sum\,\sum\limits_{i,j \in E(0.9)} I(i,j)}
\end{equation}
where $I(i,j)$ is the original image pixel index.

\item[$\clubsuit$] \textbf{Bershady--Conselice concentration index}  \citep[C$_{BC}$,][]{Bershady2000}. This parameter is defined as the ratio between the circular radii containing 80\% and 20\% of the total flux:

\begin{equation}
C_{BC}\,=\,5\,\log \frac{r_{80}}{r_{20}}
\end{equation}

\item[$\clubsuit$] \textbf{Gini coefficient} \citep[GINI,][]{Abraham2003,Lotz2004}. It is a statistic parameter based on the Lorentz curve that presents the cumulative distribution function of galaxy's pixel $i$ values:

\begin{equation}
\mathop{\mathrm{GINI}} =
{\frac{1}{|\bar{X}|n(n - 1)} \sum_{i}^{n} (2i - n - 1) |X_i|,}
\end{equation}

where $n$ is a total number of pixels in a galaxy, X$_i$ the pixel flux value, and $|\bar{X}|$ the mean over all pixel flux values. It is another type of concentration index, and usually correlates with C and C$_{BC}$. However, unlike the previous concentration indexes, it can distinguish between the galaxies with shallow light profiles and those with the light concentrated in few pixels, but outside the galaxy centre.

\item[$\clubsuit$] \textbf{Asymmetry} \citep[A,][]{Abraham1996, Conselice2000}. It measures the degree to which the light of the galaxy is rotationally symmetric. It is quantified by subtracting from the original image the galaxy rotated by 180$^\circ$ :

\begin{eqnarray}
A\,=\, \frac{1}{2} \left( \frac{\sum (|\,I(i,j)\,-\,I_{180}(i,j)\,|)}{\sum I(i,j)}\,-\right.\nonumber\\
\left.\frac{\sum(|\,B(i,j)\,-\,B_{180}(i,j)\,|)}{\sum I(i,j)} \right) ,
\end{eqnarray}

where $I$ is the flux in the original image, $B$ is the flux in the background image, while the subindex $180$ refers to the original and background images rotated by 180$^{\circ}$ around the central pixel of the galaxy.

\item[$\clubsuit$] \textbf{Smoothness or clumpiness} \citep[S,][]{Conselice2003}. Quantifies the degree of small-scale structure of the galaxy. The image is smoothed with a box of width 0.25\,$r_p$ and then subtracted from the original image:

\begin{eqnarray}
S\,=\, \frac{1}{2} \left( \frac{\sum (|\,I(i,j)\,-\,I_{S}(i,j)\,|)}{\sum I(i,j)}\,-\right.\nonumber\\
\left.\frac{\sum(|\,B(i,j)\,-\,B_{S}(i,j)\,|)}{\sum I(i,j)} \right),
\end{eqnarray}

where $I$ and $B$ are defined as in the case of $A$, while the subindex $S$ refers to the smoothed image. The resulting image provides information about possible clumpy regions.

\item[$\clubsuit$] \textbf{M$_{20}$ moment of light} \citep[][]{Lotz2004}. It is defined as a normalized second--order moment of the 20\% brightest pixels of the galaxy:

\begin{equation}
M_{20}\,=\, \log \frac{\sum\limits_{i}^{\sum f_i\,<\,0.2\,f_{tot}} M_i}{M_{tot}}
\end{equation}

where $f_i$ is the flux at the $i$ pixel, and $M_{tot}$ the total second--order moment, defined as the flux in each pixel $(x_i,y_i)$ multiplied by the squared distance to the centre of the galaxy ($x_c$ and $y_c$), and summed over all the galaxy pixels assigned by the segmentation map:

\begin{equation}
M_{tot}\,=\, \sum\limits_{i}^{n} f_i[(x_i-x_c)^2+(y_i-y_c)^2)].
\end{equation}

This parameter can trace the spatial distribution of any bright core, bar, spiral arm, and off--centre star cluster.

\end{itemize}

	\subsection{Description of the training set of local galaxies and related methodology}
    
\begin{figure}
\centering
\includegraphics[width=\hsize]{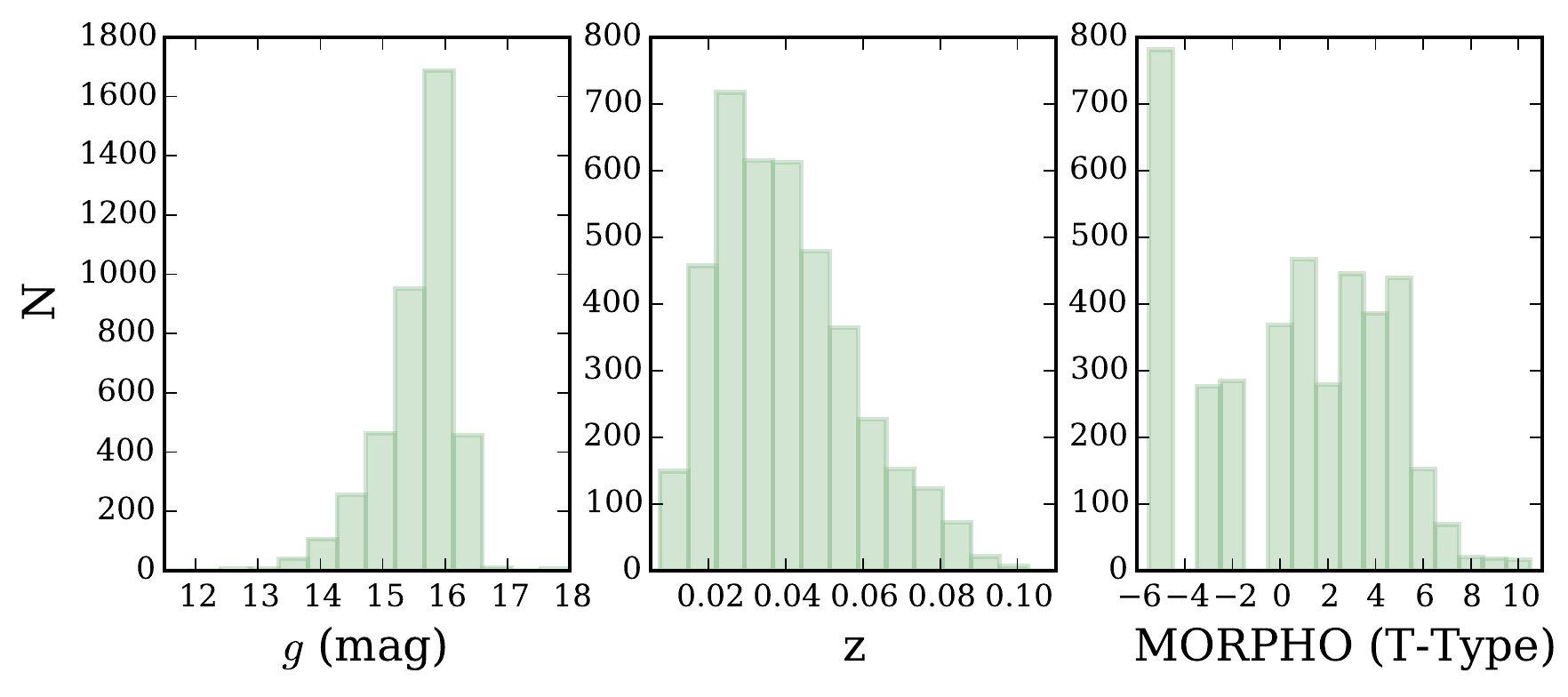}
\caption[Histogram of the local sample magnitude, redshift and T--Type]{Distributions of the $g$--band magnitude (\textit{left}), redshift (\textit{centre}), and morphological type (\textit{right}) for the  training sample of 4000 local galaxies.}
\label{fig:morphoLocal}
\end{figure}

In this work we used a sample of 4000 local galaxies classified visually by \cite{Nair2010}, using the $g$ and $r$ bands. The galaxies have redshifts distributed in the range 0.01\,$\le$\,z\,$\le$\,0.1, and were drawn from the SDSS Data Release 4 (DR4) down to an apparent extinction--corrected magnitude of $g$\,$<$\,16. For all galaxies we provided as an input information for galSVM a catalogue that includes redshift, magnitude, morphological type and half--light radius, plus original, PSF, and mask images. The build of training set of local galaxies is described in \cite{Povic2013}. The galaxies were selected randomly out of $\sim$\,14,000 sources contained in the \cite{Nair2010} catalogue, making sure that the selected sub--sample is representative in terms of general properties of the whole data set: $g$ band magnitude, redshift, $g$\,-\,$r$ colour, morphological classification, and inclination in case of the late--type galaxies. On the other hand, the number of local galaxies is selected as a compromise between the computing time and the classification accuracy, since the computing time to train the SVM through the galSVM code is totally dependent and sensitive to the size of the training data set \citep{HuertasCompany2008,HuertasCompany2009b}. Figure\,\ref{fig:morphoLocal} shows the redshift, $g$--band magnitude, and morphological classification (T--Type in \cite{Nair2010} classification) distributions of the selected local sample. As can be seen from the T--Type histogram, the selected sample spans all range of morphologies, from elliptical to irregular galaxies.

This local sample of 4000 galaxies was moved to the appropriate redshift, luminosity and image quality for the RXJ1257 field. To that purpose, galSVM requires, along with the image characteristics (pixel scale\,=\, 0.254$''$px$^{-1}$, FWHM\,=\,1.1$''$, gain\,=\,0.95, saturation level\,=\,30000), the apparent magnitude and redshift distributions of the cluster sources. For every local galaxy stamp, galSVM generates a random pair of magnitude and redshift values, with a probability distribution matching that of the cluster galaxies. Moreover, each simulated local galaxy is re--sampled with the corresponding pixel scale and convolved with the PSF to match the same spatial resolution. 

In the subsequent step we dropped the simulated galaxies into the
background of the RXJ1257 image by randomly selecting locations free
of objects (by means of the NUMBER SExtractor parameter and the segmentation
image). We then used the SExtractor mask image to eliminate nearby objects.

Finally, we measured the simulated galaxies to estimate the set of
morphological parameters listed in Sect. 3.1 and the ellipticity, as
determined by SExtractor. We used these seven parameters simultaneously
to train the SVM. The particular set of parameters used to obtain the morphological classification was tested in \citet{HuertasCompany2008,HuertasCompany2009b} and \citet{Povic2012, Povic2013}. 

\subsection{Methodology applied on the RXJ1257 galaxies}
\label{subsec:morphoClas_method_real}

\begin{figure}
\centering
\includegraphics[width=\hsize]{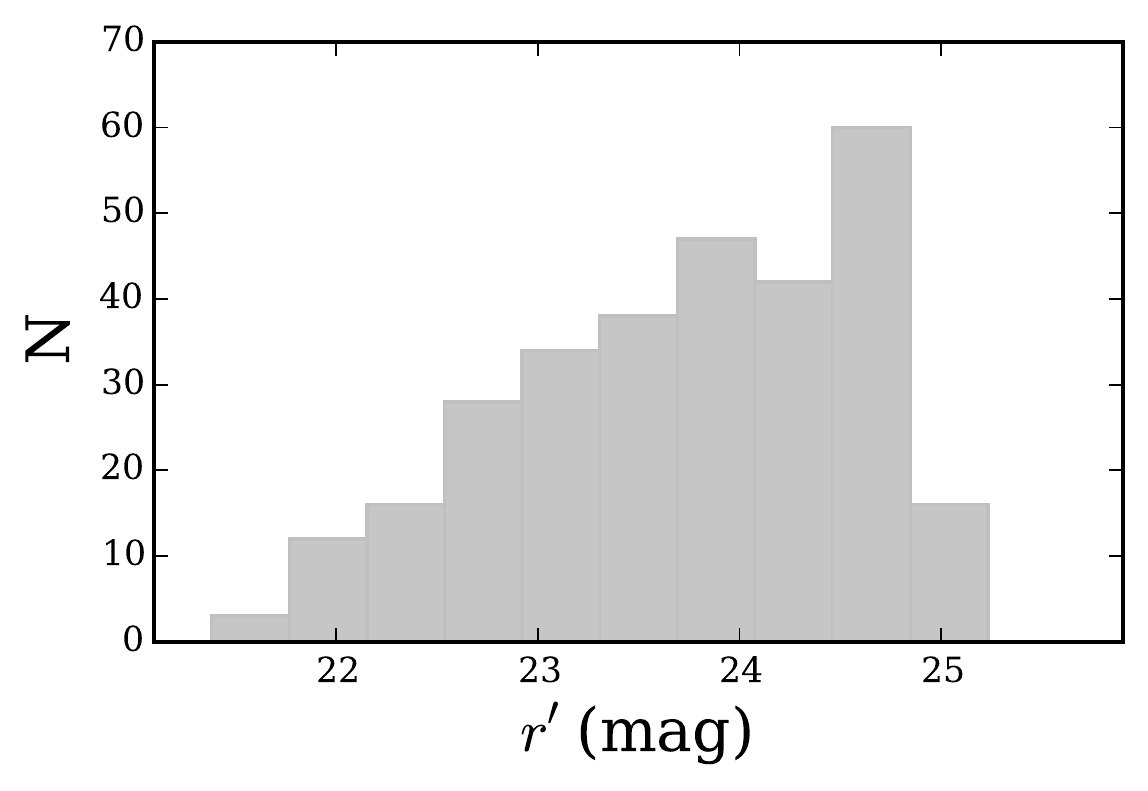}
\caption{Histogram of the $r'$ magnitude that corresponds to the full cluster sample.}
\label{fig:morphoMag}
\end{figure}

\begin{figure}
\centering
\includegraphics[width=\hsize]{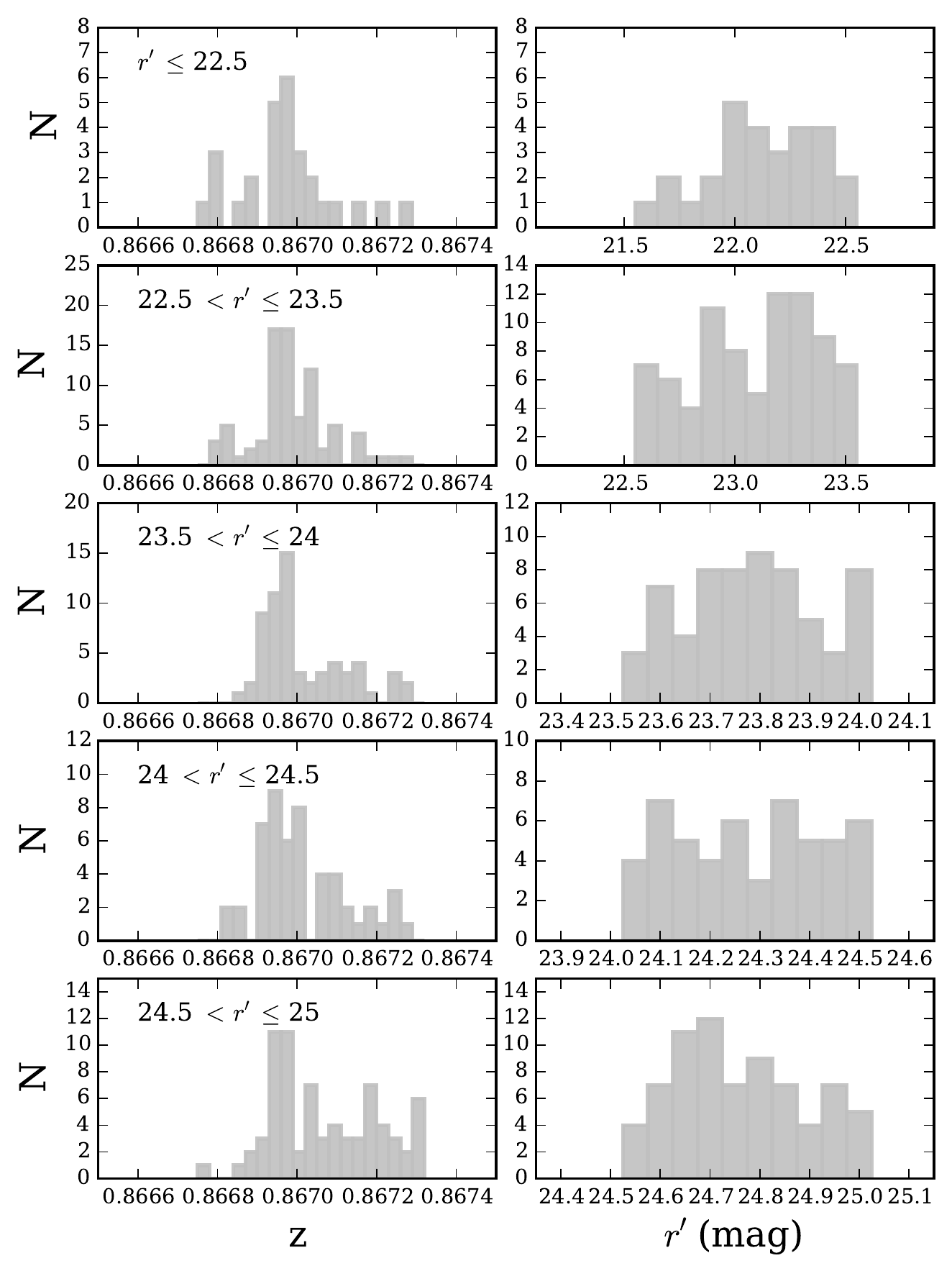}
\caption{Distributions of redshift (\textit{left}) and magnitude (\textit{right}) for each of the bins used to run the galSVM code.}
\label{fig:morphoZmag}
\end{figure}

Using galSVM code we measured the same set of seven morphological parameters of the RXJ1257 members, like in the case of simulated local galaxies. We used the $r'$--band image to perform the classification of cluster galaxies since it shows the highest signal--to--noise ratio. To take care of the k--correction effect when simulating local galaxies, SDSS $u$--band images correspond to the selected band at the redshift of our cluster. Since the filter efficiency for the SDSS u band is very poor and much lower than that for $g$, $r$, and $i$ bands \citep{Gunn1998}, and taking into account that the aim of our work is not to deal with the fine morphology, but to classify galaxies into only two broad morphological groups (early- or late-type), we used the $g$ band images instead of $u$ \citep[see][]{Povic2015temp}. Support vector--based learning machine were trained with a sample of 3000 out of 4000 initial simulated local galaxies. This number of galaxies was selected to be high, but still low enough to include different galaxies from one Monte--Carlo run to another. As a result of an empirical trade--off between computing time and accuracy tested with 10, 15 and 20 Monte--Carlo simulations, we repeated the classification of our cluster galaxies through 15 independent Monte--Carlo runs. As mentioned above, in each run we use a different combination of 3000 local galaxies out of 4000 initially selected, increasing therefore the accuracy of our classification. From each Monte--Carlo simulation we obtained the probability for every galaxy to be ET, and measured the final probability as the average of these probability values, and its error as the standard deviation of the probability distribution.

galSVM code is optimized to classify faint galaxies. This means that, if using a wide range of magnitudes, the fraction of misclassified bright objects could be significant \citep{HuertasCompany2008}. In our case, the RXJ1257 cluster sample is dominated by faint sources, as shown by the shape of the magnitude distribution in Fig.\,\ref{fig:morphoMag} where 49\% of the sources are fainter than $r'$\,=\,24. Therefore, to avoid this effect, we performed the morphological classification within five magnitude bins: m\,$\leq$\,22.5, 22.5\,$<$\,m\,$\leq$\,23.5, 23.5\,$<$\,m\,$\leq$\,24, 24\,$<$\,m\,$\leq$\,24.5, and 24.5\,$<$\,m\,$\leq$\,25. The redshift and magnitude distributions that correspond to each magnitude bin are shown in Fig.\,\ref{fig:morphoZmag}. These are the distributions that we used for simulating the local galaxy sample. We also tested the morphological classification when dealing with increasing magnitude cuts: m\,$\leq$\,22.5, m\,$\leq$\,23.5, m\,$\leq$\,24, m\,$\leq$\,24.5, and m\,$\leq$\,25, but we found that in some cases fainter objects were slightly better classified (with higher probabilities) using smaller magnitude bins.

\subsection{Estimation of errors of the morphological parameters}
\label{subsec:morphoClas_errors}

\begin{figure*}
\centering
\includegraphics[width=0.8\hsize]{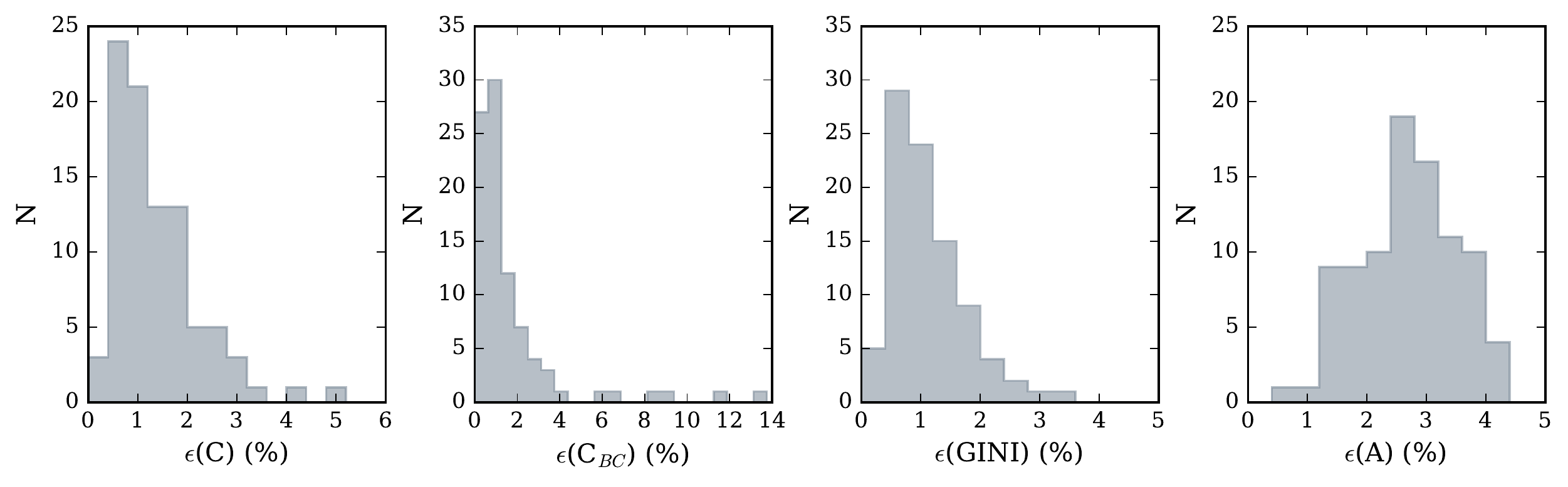}
\caption{Distributions of the percentage errors associated to C, C$_{BC}$, GINI, and A morphological parameters.}
\label{fig:morphoErr}
\end{figure*}

The galSVM code does not estimate uncertainties of the measured morphological parameters. The only output error estimate is the standard deviation of the probabilities determined in each Monte Carlo simulation. In order to get an idea  of the errors associated to the morphological parameters that have been used to perform the classification with the galSVM code, we proceeded as follows: (i) we estimated the $r'$--band image noise; (ii) we generated 100 mock images by randomly varying every pixel counts within a Gaussian of sigma value equal to the image noise; (iii) we measured the morphological parameters used in each mock image; (iv) for each galaxy we obtained a distribution of each morphological parameter, and we associated its standard deviation with its error. Figure\,\ref{fig:morphoErr} shows the distributions of percentage errors for the C, C$_{BC}$, GINI, and A morphological parameters obtained for the cluster galaxies morphologically classified. We observe in this figure that, with the exception of the error associated to the C$_{BC}$  perameter of a set of six galaxies, the error values are lower than 5\%, being around  1\% for the concentration indexes and around  3\% for the asymmetry. The M20 parameter was omitted in this figure since it remains invariant under image noise variations.

\section{Morphological classification: results}
\label{sec:morphoClas_results}

	Before discussing the properties of the classified galaxies, it is necessary to define the limits within which we could consider the morphological classification to be reliable. The probability that a galaxy is ET (or LT) represents a measurement of the accuracy of the morphological classification. It has been shown that there is a clear correlation between the $p_{ET}$ probability threshold and the number of correct identifications: accuracy clearly increases when the considered probability is higher. \citet{HuertasCompany2009a} compared WIRCam data at the Canada France Hawaii Telescope with HST/ACS observations, and found that objects with $p_{ET}$ probabilities between 0.5 and 0.6 have mean accuracy of around 58\%, while objects with probabilities greater than 0.8 are classified as ET with accuracy of nearly 90\%. Due to the unavailability of higher--resolution and comparable morphological classifications, in order to establish the probability boundaries under which our morphology is reliable, we analysed probability distributions and standard diagnostic diagrams. We also tested the physical properties of classified sources. 
    
\begin{figure}
\centering
\includegraphics[width=\hsize]{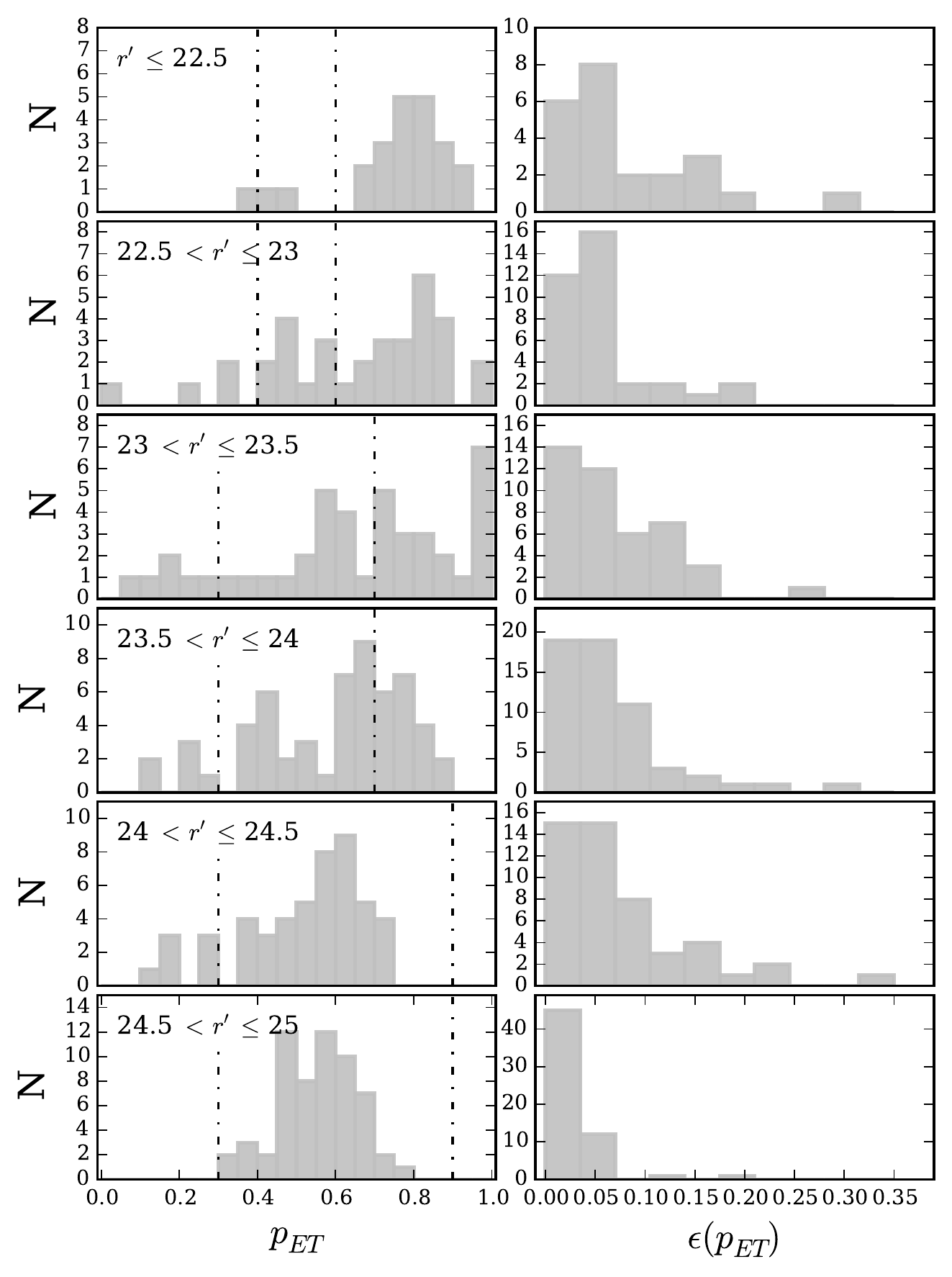}
\caption[Morphological probability distribution]{Distribution of the $p_{ET}$ (\textit{left}) and of their errors (\textit{right}) for each analysed bin of magnitude. The dot-dashed vertical lines mark the final probability boundaries we have considered for the separation between morphological types, ET and LT galaxies, as indicated in Tab.\,\ref{tab:morphoprob}.}
\label{fig:morphoProb}
\end{figure}

Figure\,\ref{fig:morphoProb} shows the obtained average probability distributions and their errors in six analysed magnitude bins. As already mentioned above, the errors were estimated as the standard deviation of the probabilities obtained in the 15 Monte--Carlo simulations. In this figure, two trends can be observed: on the one hand, the brightest galaxies are mainly classified as ETs, as expected in a cluster, while those classified as LTs show fainter magnitudes; on the other hand, towards higher magnitude bins to distinguish between both types becomes more complicated, resulting in a peak of probability distribution around 0.5. For all magnitude bins, the probability errors are kept mostly below 0.1. 

\begin{figure*}[h!]
\centering
\includegraphics[width=0.45\hsize]{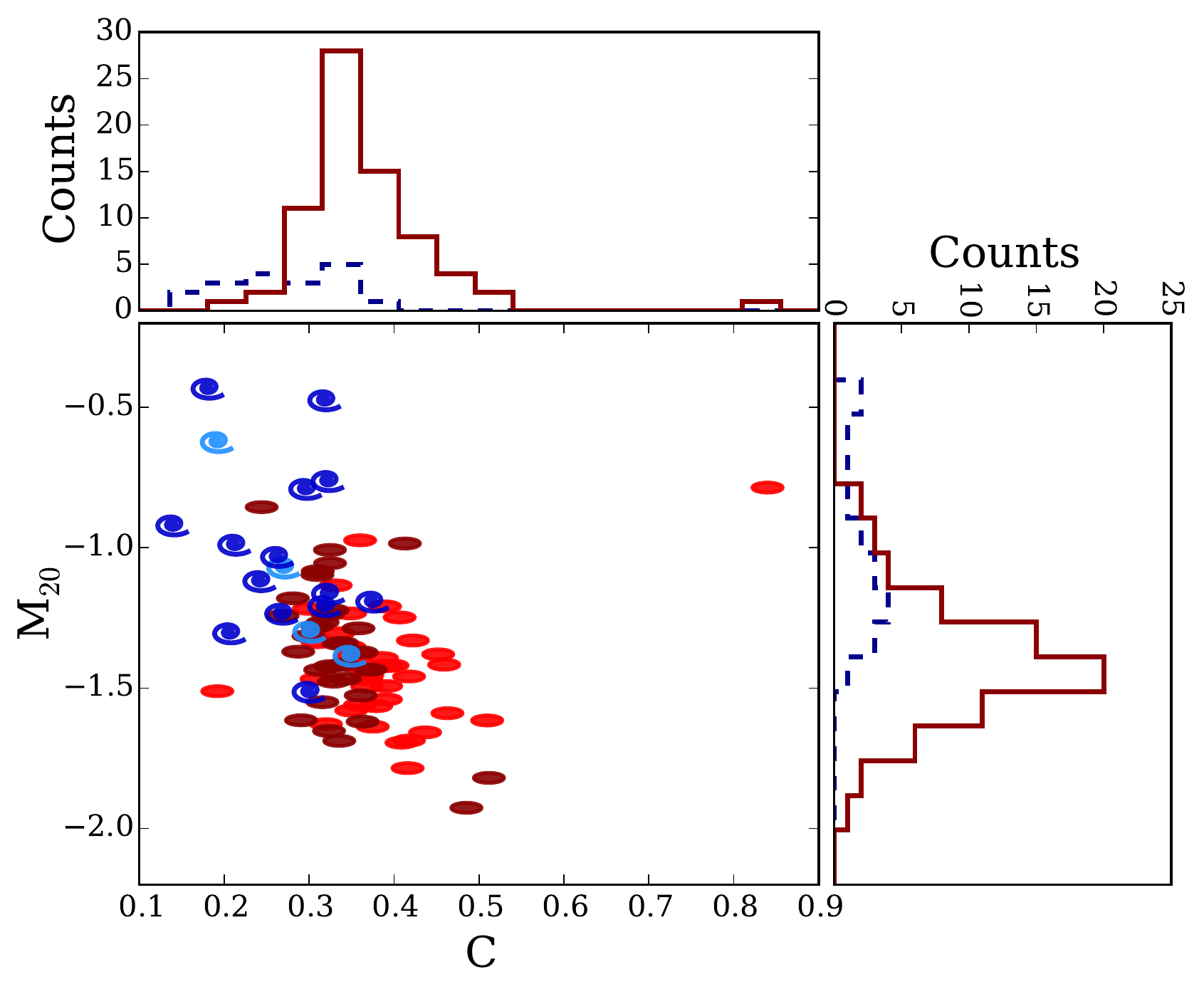}
\includegraphics[width=0.45\hsize]{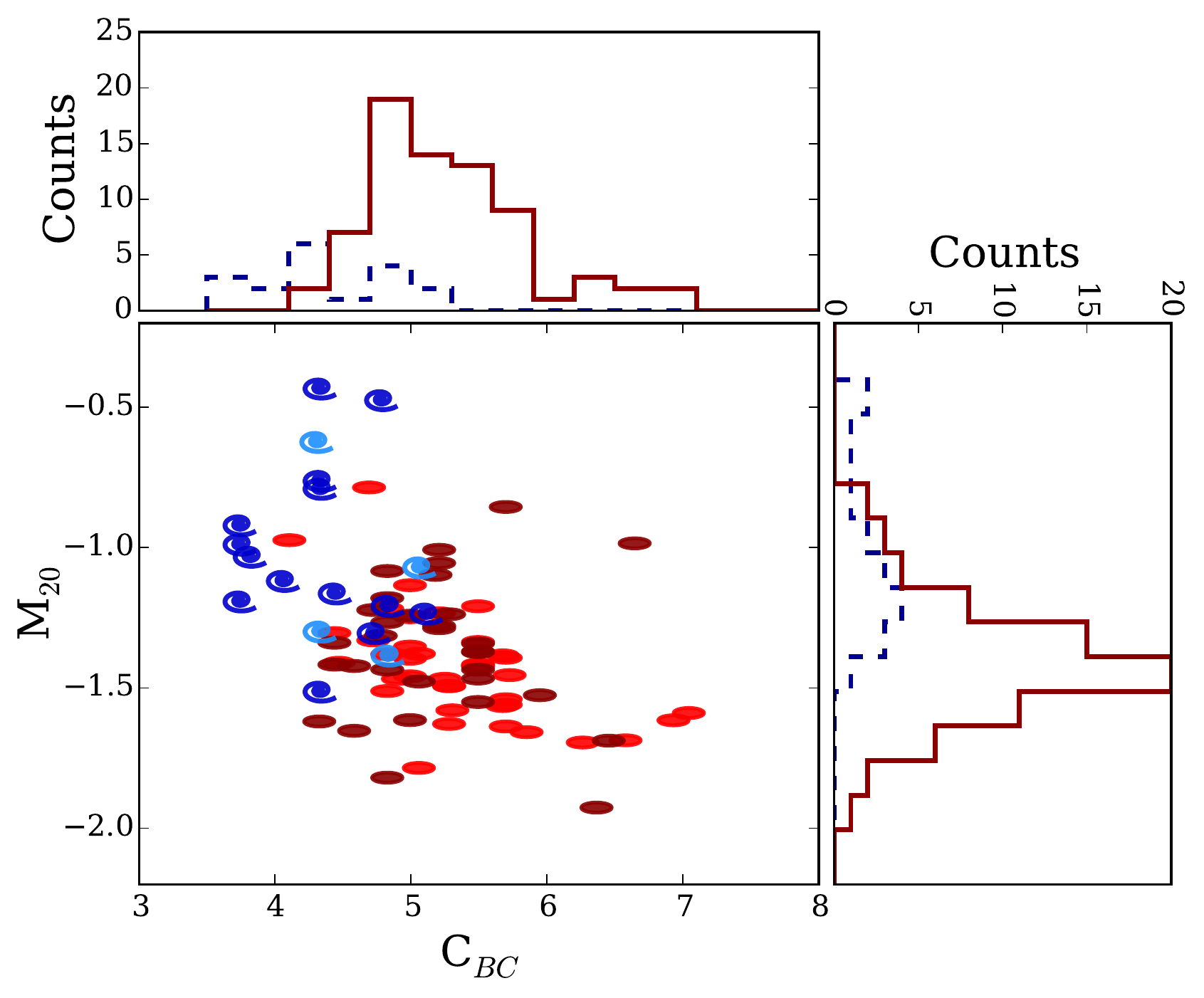}
\includegraphics[width=0.45\hsize]{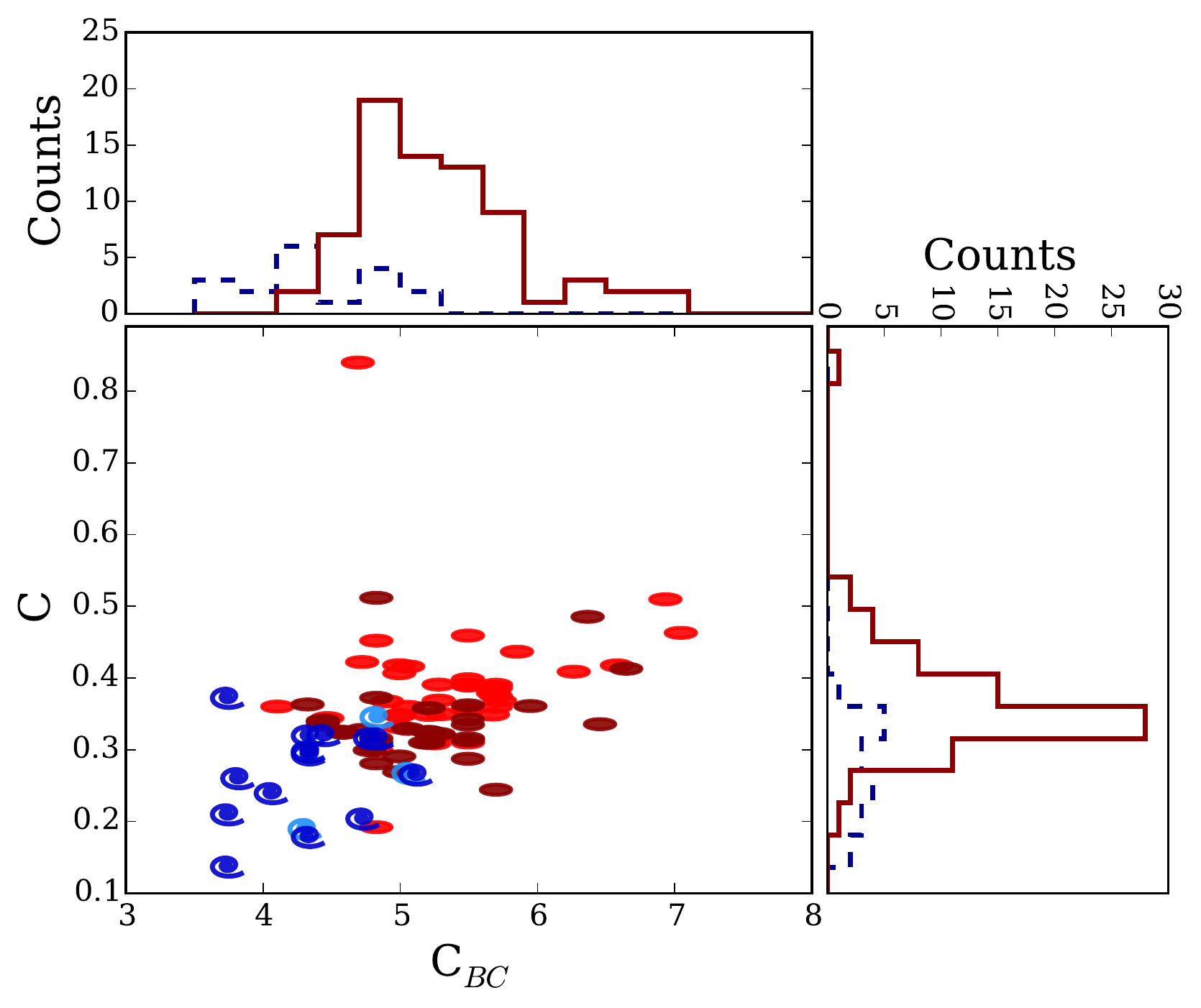}
\includegraphics[width=0.45\hsize]{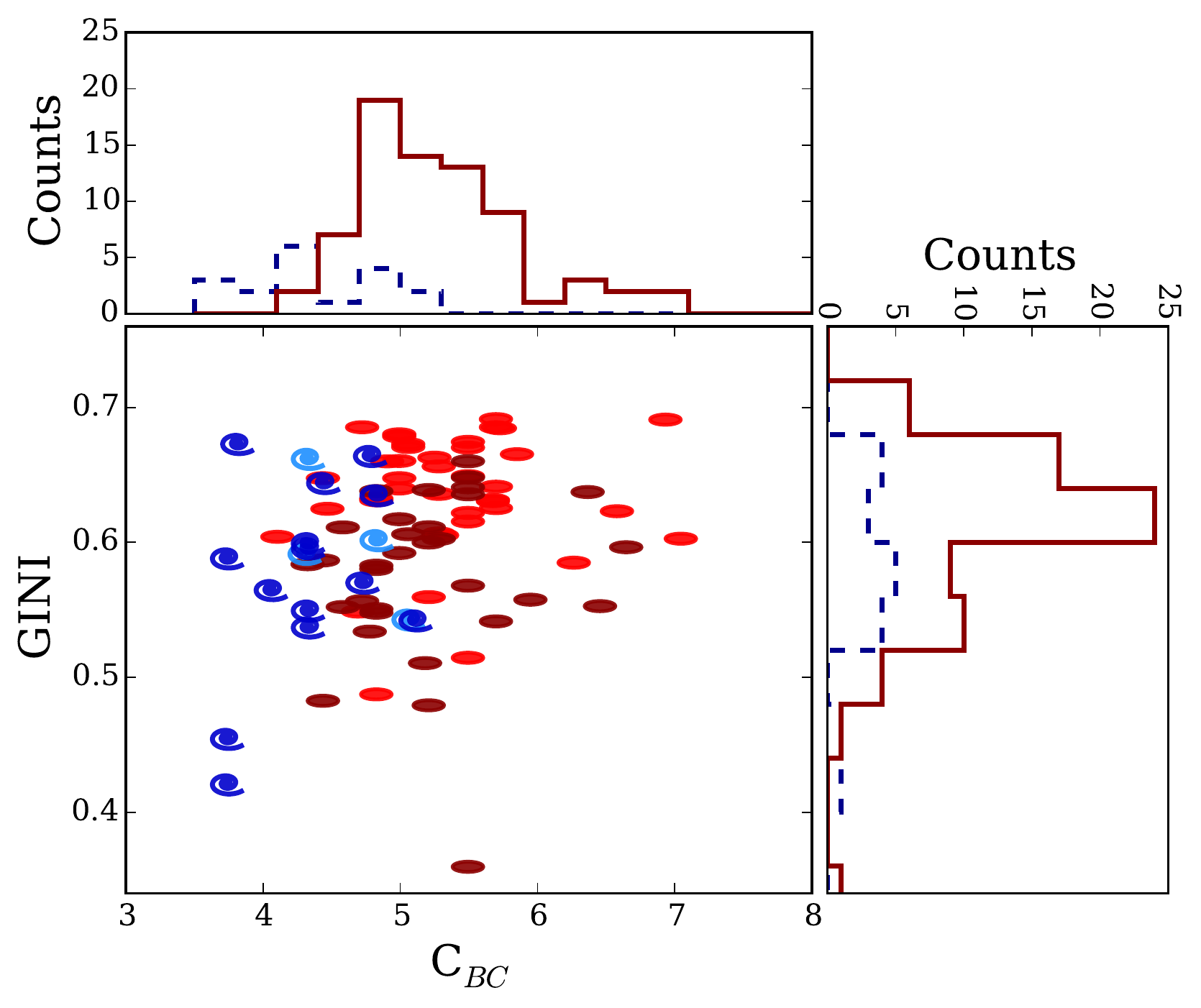}
\includegraphics[width=0.45\hsize]{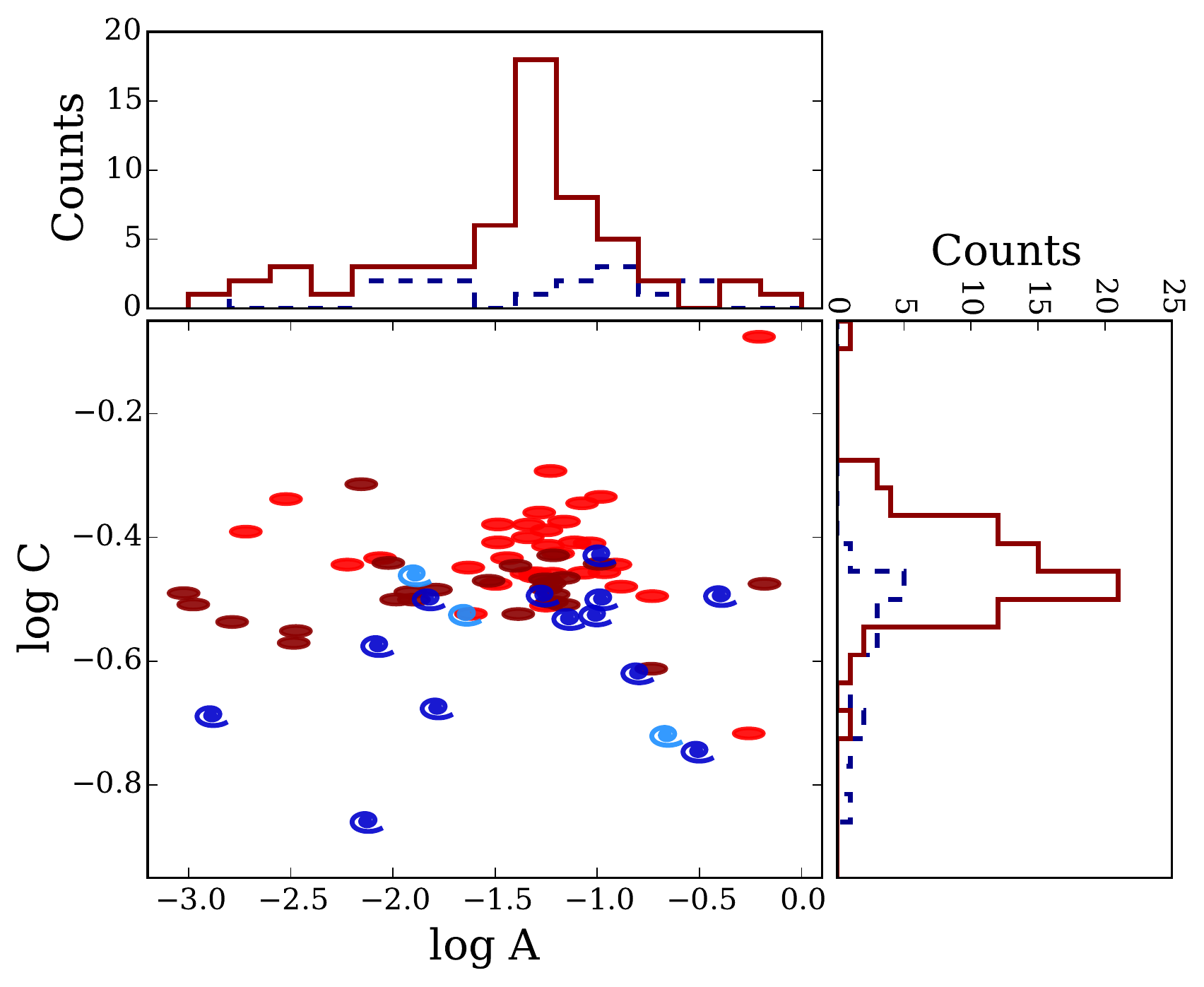}
\caption[Morphological diagnostic diagrams]{Morphological diagnostic diagrams representing the following relations: M$_{20}$ (moment of light) vs C (Abraham concentration index) (\textit{top left}), M$_{20}$  vs C$_{BC}$ (Bershady--Conselice concentration index) (\textit{top right}), C vs C$_{BC}$ (\textit{centre left}), GINI (Gini coefficient) vs C$_{BC}$ (\textit{centre right}), and $\log$\,C vs $\log$\,A (asymmetry) (\textit{bottom}). Blue spirals represent galaxies classified as LT, and red ellipses as ET, while light and dark colours refers to galaxies fainter and brighter than 23\,mag in $r'$, respectively. Typical errors of all these parameters are comparable to or lower than the symbols size. Histograms show the morphological parameter distribution for each type, red solid line is for ET galaxies and blue dashed one for LT galaxies.}
\label{fig:morphoDia}
\end{figure*}

We tested a set of probability limits for each magnitude bin, applying more restrictive values for fainter magnitudes, and analysed therefore standard morphological, colour-magnitude, and colour-stellar mass diagrams. When defining the probability thresholds we took into account the probability error, i.e. a galaxy should satisfy the condition $p_{ET}$\,-\,$\epsilon(p_{ET})$\,$\geq$\,$p_{th}^{ET}$ to be classified as a trustful ET galaxy, where $p_{th}^{ET}$ is the probability threshold to classify ET galaxies in a given magnitude bin, while $\epsilon(p_{ET})$ is the corresponding error. Figure\,\ref{fig:morphoDia} shows five of the standard morphological diagnostic diagrams. None of these diagrams represents smoothness since this parameter is extremly sensitive to spatial resolution, data depth, and noise, as showed in \cite{Povic2015temp}, and for most of our galaxies we obtained non-valid values. As shown in Fig.\,\ref{fig:morphoDia}, only using M$_{20}$, C$_{BC}$, and C parameters one can roughly classify ET and LT galaxies. 

In general, the individual morphological diagnostic diagrams do not segregate clearly between both types. Nevertheless, for the LT galaxies, we observed that a probability threshold of 0.3 for the faintest galaxies ($r'$\,$>$\,23\,mag, represented in dark blue colour in Fig.\,\ref{fig:morphoDia}) is acceptable to place the major number of galaxies in the expected position in the morphological diagnostic diagrams \citep[e.g.][]{Abraham1996,Lotz2004}. Therefore, we selected the LT sample including the galaxies that satisfied $p_{ET}+\varepsilon(p_{ET})$\,$\leq$\,0.4 for the brightest galaxies, i.e. with magnitude $r'$\,$\leq$\,23\,mag, and $p_{ET}+\varepsilon(p_{ET})$\,$\leq$\,0.3 for the fainter ones. For the ET sample, the brightest galaxies with $r'$\,$\leq$\,23\,mag were clearly classified when $p_{ET}-\varepsilon(p_{ET})$\,$\geq$\,0.6, as showed by the probability distributions and the morphological diagnostic diagrams (see Figs.\,\ref{fig:morphoProb} and \ref{fig:morphoDia}). In contrast, we observed that fainter galaxies classified as ET tend to blend into the LT region in the morphological diagnostic diagrams. For faint galaxies the information from the galactic disc can be easily lost, increasing the number of ET misclassified galaxies due to contamination of LT ones. Therefore, we established two more restrictive probability thresholds: $p_{ET}-\varepsilon(p_{ET})$\,$\geq$\,0.7 for 23\,$<$\,$r'$\,$\geq$\,24, and $p_{ET}-\varepsilon(p_{ET})$\,$\geq$\,0.9 for the faintest galaxies. The probability classification thresholds for both ET and LT types are summarized in Table\,\ref{tab:morphoprob}, and plotted as dot--dashed vertical in Fig\,\ref{fig:morphoProb}. In total, we obtained the reliable classification for 30\% of all cluster members.

		\begin{table}
	    \caption{galSVM probability threshold for selecting ET and LT galaxies in the analysed $r'$ magnitude bins.}
	    \label{tab:morphoprob}
		\begin{center}
		\begin{tabular}{c c c}
		\hline\hline
		\multirow{2}{*}{$r'$ bin} & {ET} & {LT} \\
		 & {$p_{ET}-\varepsilon(p_{ET})$} & {$p_{ET}+\varepsilon(p_{ET})$} \\
		\hline
        21.5--22.5 & $\geq$\,0.6 & $\leq$\,0.4 \\
        22.5--23.0 & $\geq$\,0.6 & $\leq$\,0.4 \\
        23.0--23.5 & $\geq$\,0.7 & $\leq$\,0.3 \\
        23.5--24.0 & $\geq$\,0.7 & $\leq$\,0.3 \\
        24.0--24.5 & $\geq$\,0.9 & $\leq$\,0.3 \\
        24.5--25.0 & $\geq$\,0.9 & $\leq$\,0.3 \\
        \hline
		\end{tabular}
        \begin{flushleft}
        \end{flushleft}
		\end{center}
		\end{table}		

\begin{figure*}
\centering
\includegraphics[width=\hsize]{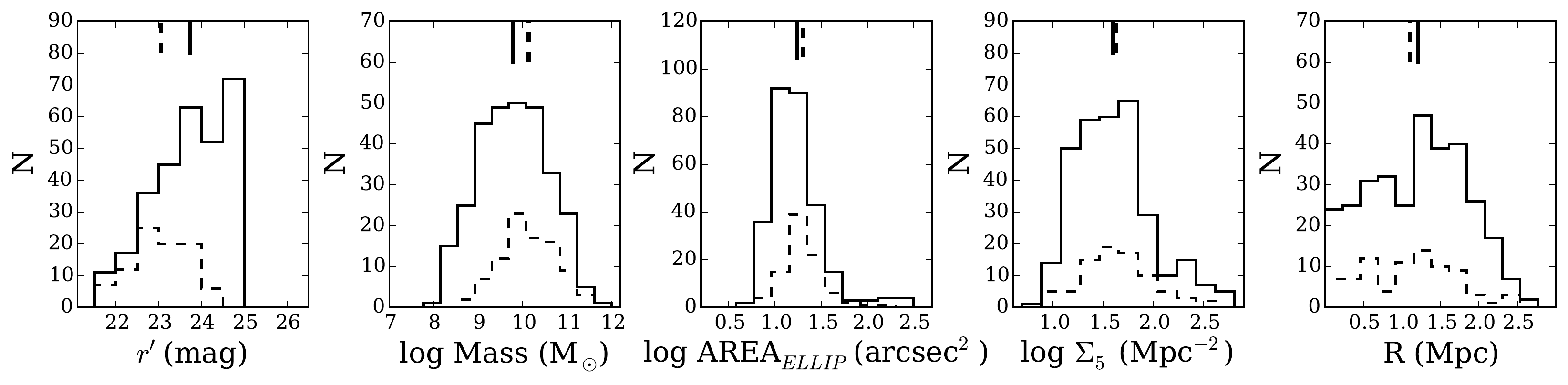}
\caption{Distributions of the $r'$ magnitude, stellar mass, size, local galaxy density, and cluster--centric distance. The solid line represents the full cluster sample and the dashed line is the morphologically classified sub--sample. Vertical lines indicate the average value of each distribution.}
\label{fig:morphoBias}
\end{figure*}

Figure\,\ref{fig:morphoBias} shows the distributions of $r'$ magnitude, stellar mass, size, local galaxy density, and distance to the cluster centre of the whole cluster sample and of the classified one (summarized in Table\,\ref{tab:morphoNum}). The size was estimated as an area covered by the ellipse used by SExtractor to extract the source. The local galaxy density $\Sigma_{5}$ was defined as the inverse of the area that comprises the five nearest neighbours \citep{Tanaka2005}. Our classification is incomplete towards fainter galaxies since all galaxies fainter than $r'$\,=\,24.5 failed to be reliably classified (see Table\,\ref{tab:morphoprob} and \ref{tab:morphoNum}). Consequently, our sample of classified cluster members is incomplete in stellar mass, with less massive galaxies remaining unclassified. Nevertheless, both the whole galaxy sample and the classified subsample appear to be consistent in terms of other properties, such as size, age, and morphological parameters. In relation with the environment, neither the local galaxy density nor the cluster--centric distance distributions show differences between the whole and classified samples. In fact, applying the Kolmogorov--Smirnov \citep[e.g.][]{Fasano1987} statistics we obtained high p--values (56 and 36\% for local density and cluster--centric distance, respectively) which confirmed the hypothesis that both distribution are drawn from the same parent population.

\begin{figure}
\centering
\includegraphics[width=\hsize]{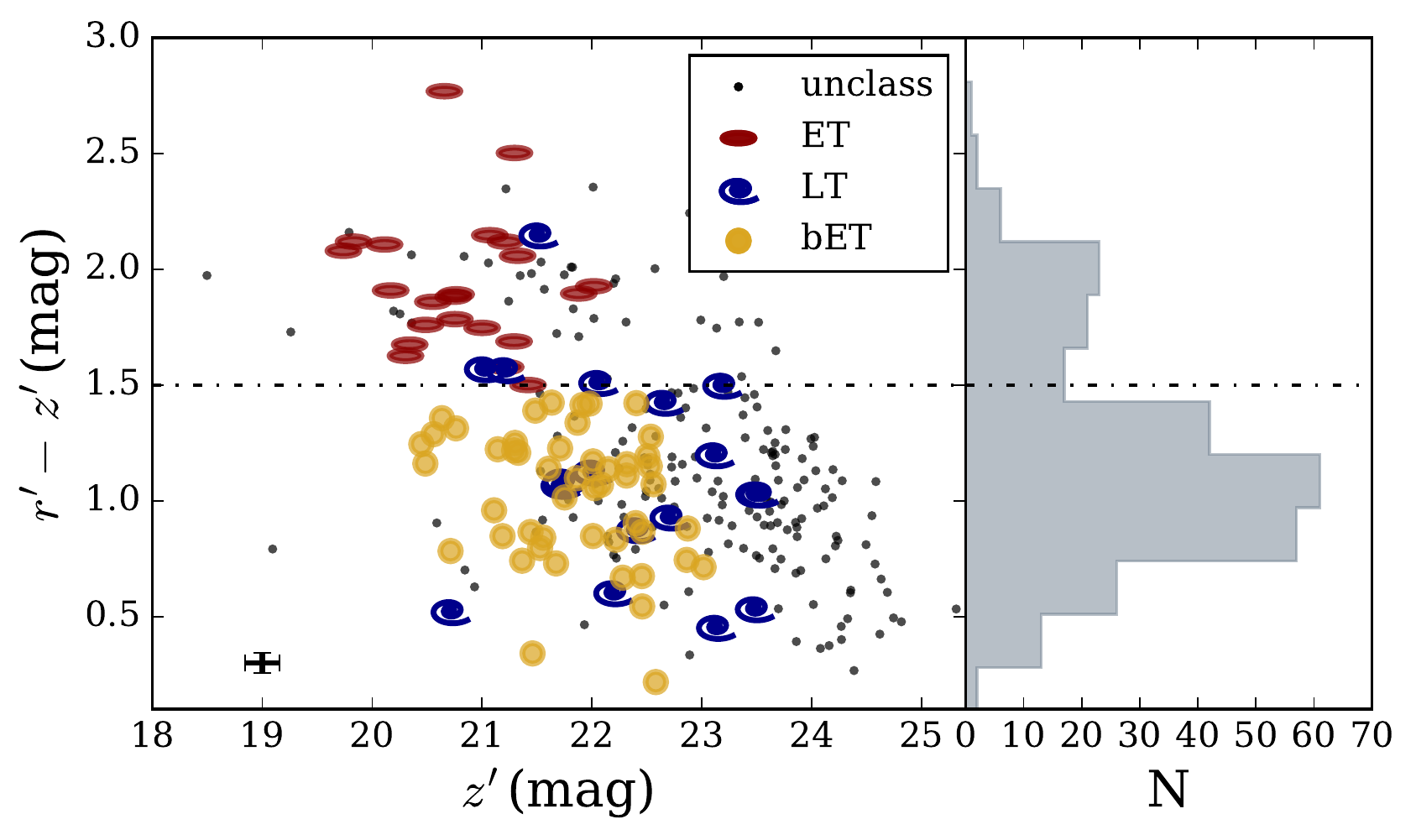}\\
\caption{CMD of RXJ1257 (\textit{left}) and $r'$\,-\,$z'$ colour distribution (\textit{right}). Symbols are the same as in Fig.\,\ref{fig:morphoDia}. Blue spirals represent galaxies classified as LT, red ellipses as ET, and yellow circles are ``blue'' ET galaxies. The horizontal dot--dashed line at $r'$\,-\,$z'$\,=\,1.5\,mag is used to separate between the red and blue populations (and, consequently, to differentiate between ``standard'' and ``blue'' ET galaxies). Small black dots are morphologically unclassified cluster members. The typical error bar is shown in the lower left conner of the plot. In the left panel, the histogram shows the whole cluster sample.}
\label{fig:morphoColMag}
\end{figure}

Once the probability boundaries were established, we were able to study the physical properties of morphologically classified cluster sample. First, we analysed their optical colours through the $r'$\,-\,$z'$ vs. $z'$ colour--magnitude diagram (CMD), shown in Fig.\,\ref{fig:morphoColMag}. We observed in the Fig.\,\ref{fig:morphoColMag} a large fraction ($\sim$64\%) of the ET population that falls in the blue region. This area is expected to be mainly populated by LT galaxies (at the 70-80\% level), even though ET galaxies with recent star formation can also be present \citep[e.g.][]{Deng2009,Schawinski2009,HuertasCompany2010}. In our case, 51\% of the morphologically classified galaxies showed this behaviour. Due to the large number of galaxies classified as ET that presented this unusual blue colour we decided to make a third morphological group and analyse its properties and dependencies separately. On the basis of the $r'$\,-\,$z'$ colour bimodal distribution, we draw the line between ``standard'' and ``blue'' ET galaxies at $r'$\,-\,$z'$\,=\,1.5\,mag (as shown in Fig.\,\ref{fig:morphoColMag} by the horizontal dot--dashed line). This threshold was determined as the $\mu$\,-\,2$\sigma$ value which corresponds to the Gaussian function fitted to the red part of the $r'$\,-\,$z'$ distribution. Besides these ``blue'' ET galaxies, we observed in this diagram that at least three LT galaxies fall into the red region. The presence of such red LT galaxies in clusters is not surprising, since they could be transition objects, from blue LT to red ET galaxies \citep{Wolf2009,Vulcani2015}. The red colour of this LT galaxy could also be caused by the presence of dust or by a high angle of inclination.
Table\,\ref{tab:morphoNum} shows the number of galaxies included in each morphological group and the percentage of well--classified galaxies relative to the total number of cluster members, overall and per magnitude bin.

		\begin{table}
	    \caption{Number of LT, ET and blue ET galaxies in each $r'$ magnitude bin, including the percentage of morphologically classified galaxies.}
	    \label{tab:morphoNum}
		\begin{center}
		\begin{tabular}{c c c c c}
		\hline\hline
		$r'$ bin & {ET} & {LT} & {blue ET} & \% Classified\\
		\hline
        21.5--22.5 & 8 & 0 & 11 & 68 \\
        22.5--23.0 & 6 & 4 & 15 & 69 \\
        23.0--23.5 & 5 & 3 & 12 & 44 \\
        23.5--24.0 & 3 & 5 & 12 & 32 \\
        24.0--24.5 & 0 & 6 & 0 & 12 \\
        24.5--25.0 & 0 & 0 & 0 & 0 \\
        \hline    
        Total & 22 & 18 & 50 & 30 \\
        \hline
		\end{tabular}
        \begin{flushleft}
        \end{flushleft}
		\end{center}
		\end{table}


\section{Physical properties of different morphological types}
\label{sec:properties}

In this section we investigate the properties of the three categories of galaxies we have identified and consider possible explanations for the unexpected population of blue ET galaxies

\begin{figure*}
\centering
\includegraphics[width=\hsize]{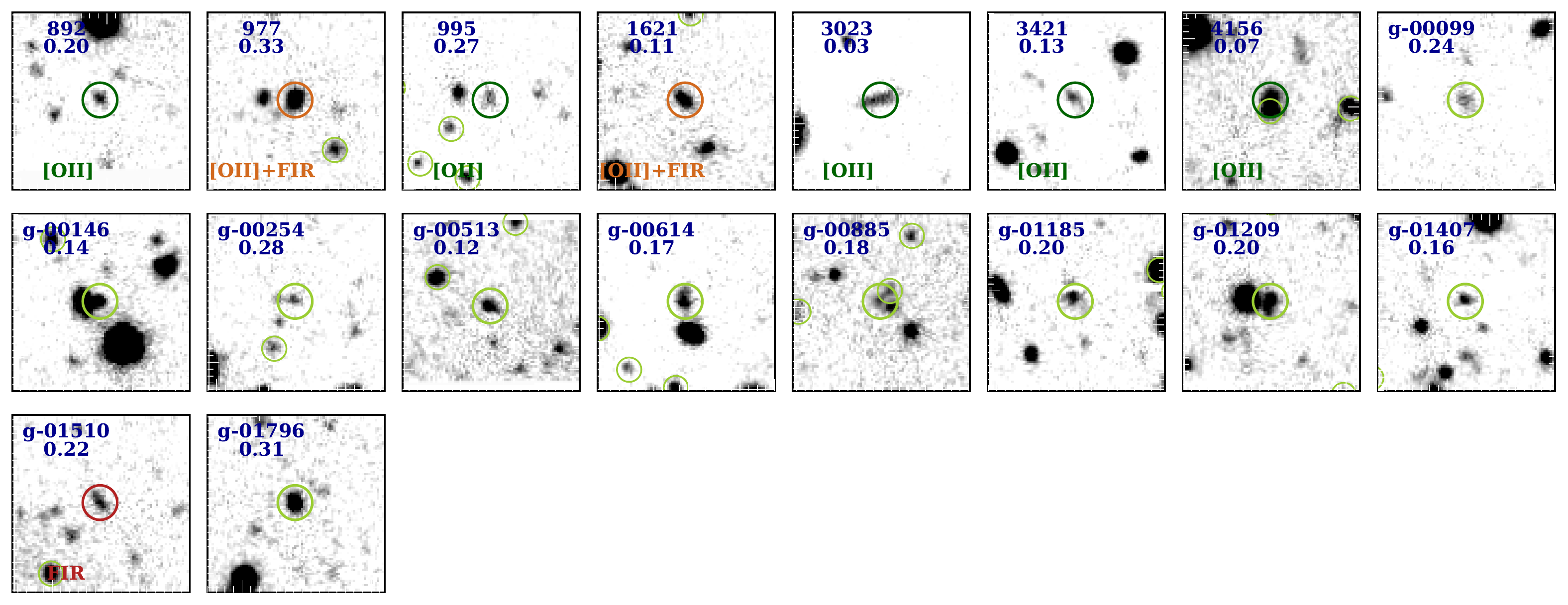}
\caption{Thumbnails of the galaxies classified as LT, according to the probability thresholds shown in Table\,\ref{tab:morphoprob}. Upper labels are the identification name and the probability value, and lower label is, when detected, the emission type. Green circles indicate galaxies considered as cluster members. Central circles mark the classified galaxy, with the colour indicating the type of emission (dark green, red, and orange for \oii, FIR, and both, respectively). Each thumbnail is 10\,arcsec$^2$, and the colour scale is equal in every image.}
\label{fig:morphoThumbLT}
\end{figure*}

\begin{figure*}
\centering
\includegraphics[width=\hsize]{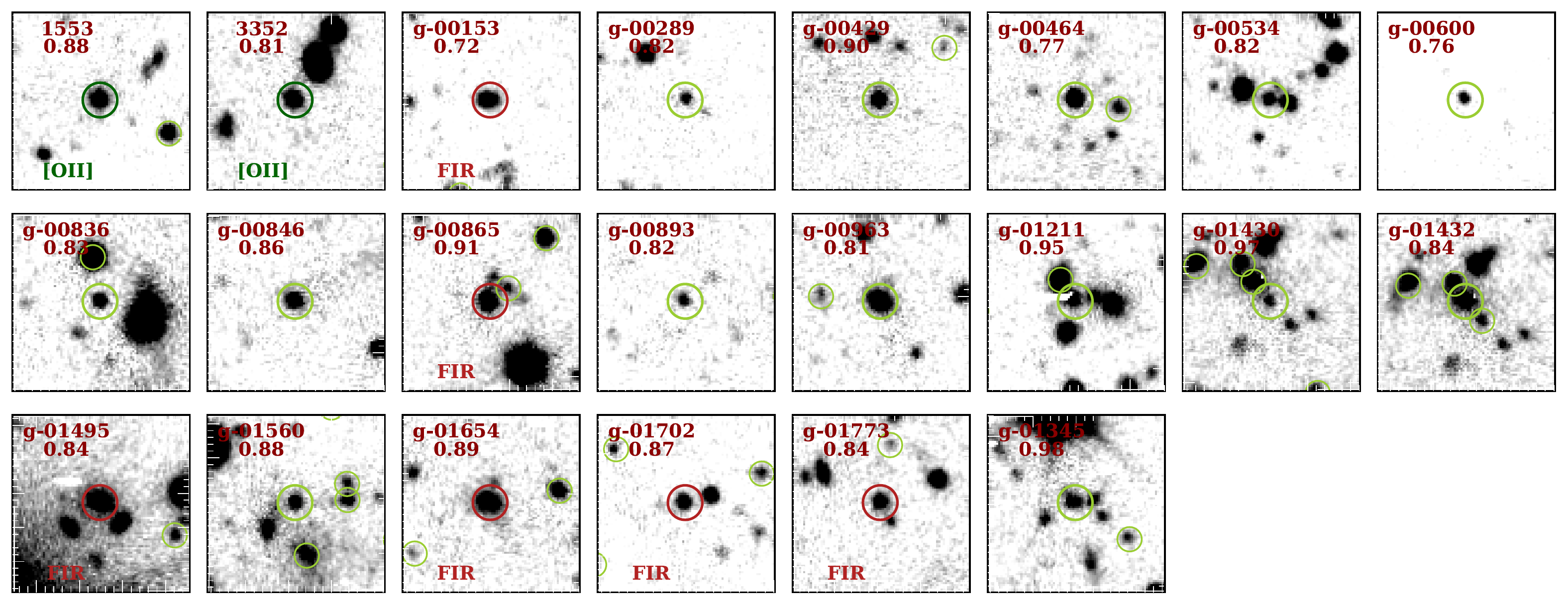}
\caption{Thumbnails of the galaxies classified as ET, according to the probability thresholds shown in Table\,\ref{tab:morphoprob} and the $g'$\,-\,$z'$ colour limit imposed for the ET. Labels and circles indicate the same as in Fig.\,\ref{fig:morphoThumbLT}.}
\label{fig:morphoThumbET}
\end{figure*}

\begin{figure*}
\centering
\includegraphics[width=\hsize]{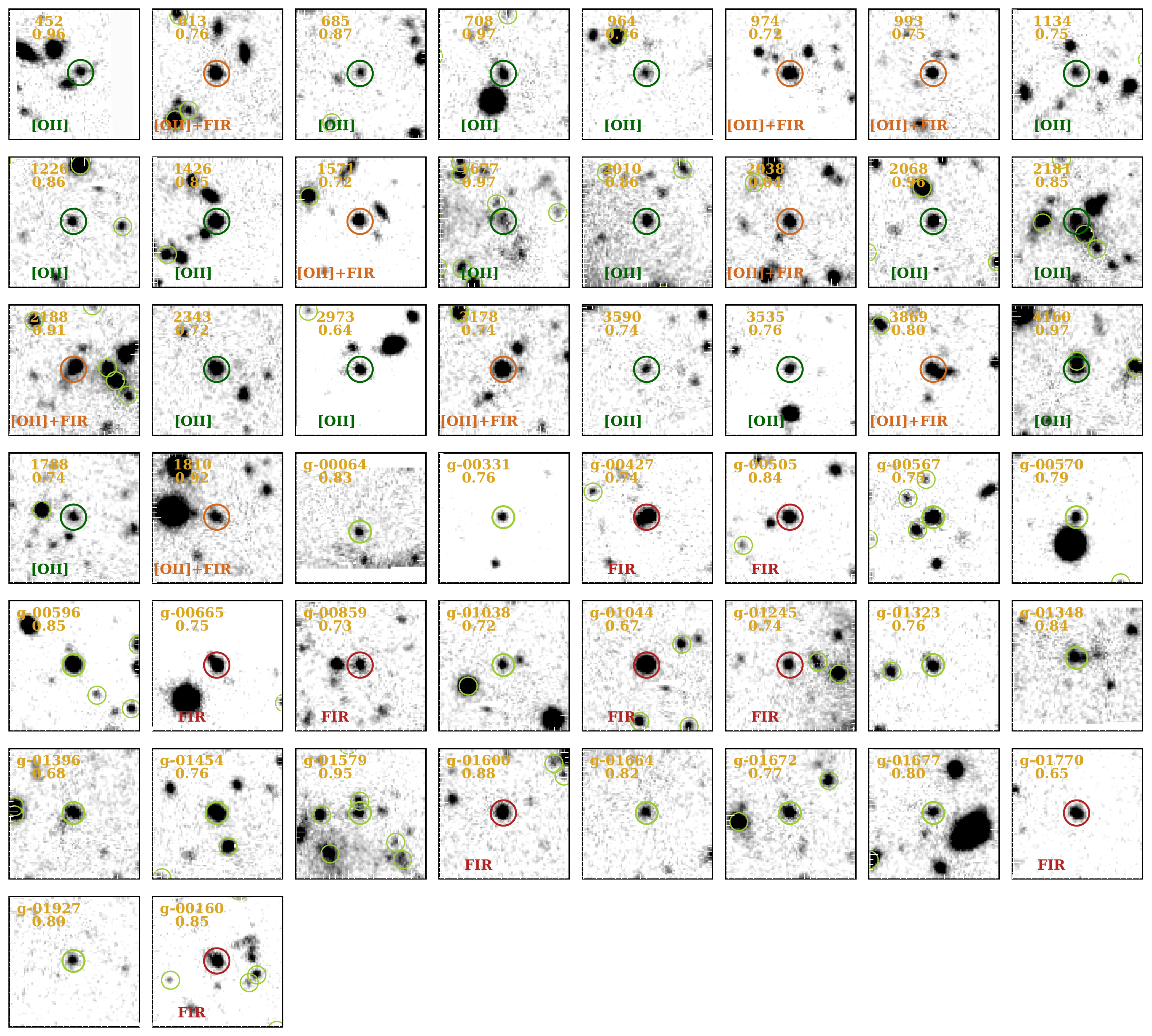}
\caption{Thumbnails of the galaxies classified as ET according to the probability thresholds (see Table\,\ref{tab:morphoprob}), but that are bluer than the $g'$\,-\,$z'$ colour limit considered for the ET. Labels and circles indicate the same as in Fig.\,\ref{fig:morphoThumbLT}.}
\label{fig:morphoThumbBlue}
\end{figure*}

First, we analysed its appearance in the $r'$--band, which is the image used for the classification. In Figs.\,\ref{fig:morphoThumbLT},\ref{fig:morphoThumbET}, and \ref{fig:morphoThumbBlue}, we show the thumbnails of the sources included in the three samples. As we could expect, LT galaxies seem more diffuse, faint, and elongated when compared with the rest. Two of them (g--00146 and g--01209) have blend problems due to a nearby bright source, although both have $p_{ET}$\,$\leq$\,0.2. Another two sources (4156 and g--00885) that show a very close companion might be mergers, since in the two cases both galaxies belong to the cluster. ET galaxies are, in general, more rounded and with the light uniformly distributed. Most of them, 17 from 24, appeared as isolated galaxies, but, in contrast, at least five of them (g--00865, g--01211, g--01430, g--01432, and g--01560) are found in apparently dense environments, with many nearby galaxy cluster members. The remaining two ETs (g--00534 and g--01495) appear very close to other sources not classified as cluster members. In the third sample of blue ET galaxies, there are a variety of galaxies: clearly faint and diffuse (e.g. 1677), bright and very rounded (e.g. g--00505), slightly elongated (e.g. 708), blended or merger (e.g. 1160), isolated (e.g. g--01664), or inside a dense environment (e.g. 2181). Nevertheless, compared with the ET and LT samples, most of the blue ET galaxies are visually more similar to the ET ones.

\begin{figure}
\centering
\includegraphics[width=\hsize]{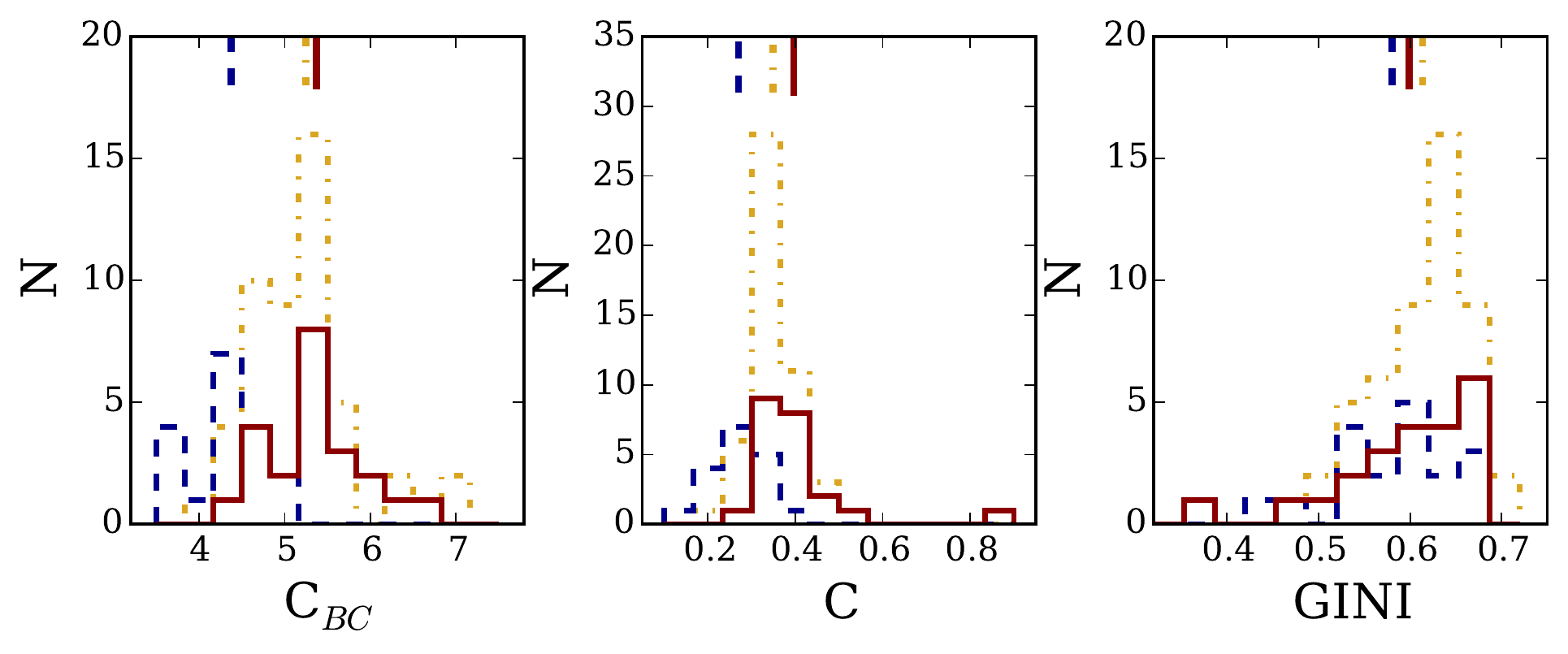}
\caption{Distributions of the morphological concentration parameters (C, C$_{BC}$, GINI) for the three morphological samples. The LT sample is represented by the dark blue dashed line, the ET sample by the red solid line, and the ``blue'' ET sample by the yellow dash--dotted line. Top vertical lines indicates the mean values of each distribution.}
\label{fig:morphoParamHist}
\end{figure}

Figure\,\ref{fig:morphoParamHist} shows the distribution of three concentration indexes (C$_{BC}$, C, and GINI) for LT, ET, and ``blue'' ET galaxies. These parameters are the most stable against the effects of survey depth, spatial resolution, and noise, and therefore the most reliable indicators of the morphology, especially when using the ground-based data \citet{Povic2015temp}. In general, ``blue'' ET galaxies are as concentrated as the standard, ``red'' ETs, being both types more concentrated than the LT sample. This is especially clear in the case of C$_{BC}$ and C parameters, while in the case of GINI the differences are less significant.

\begin{figure}
\centering
\includegraphics[width=\hsize]{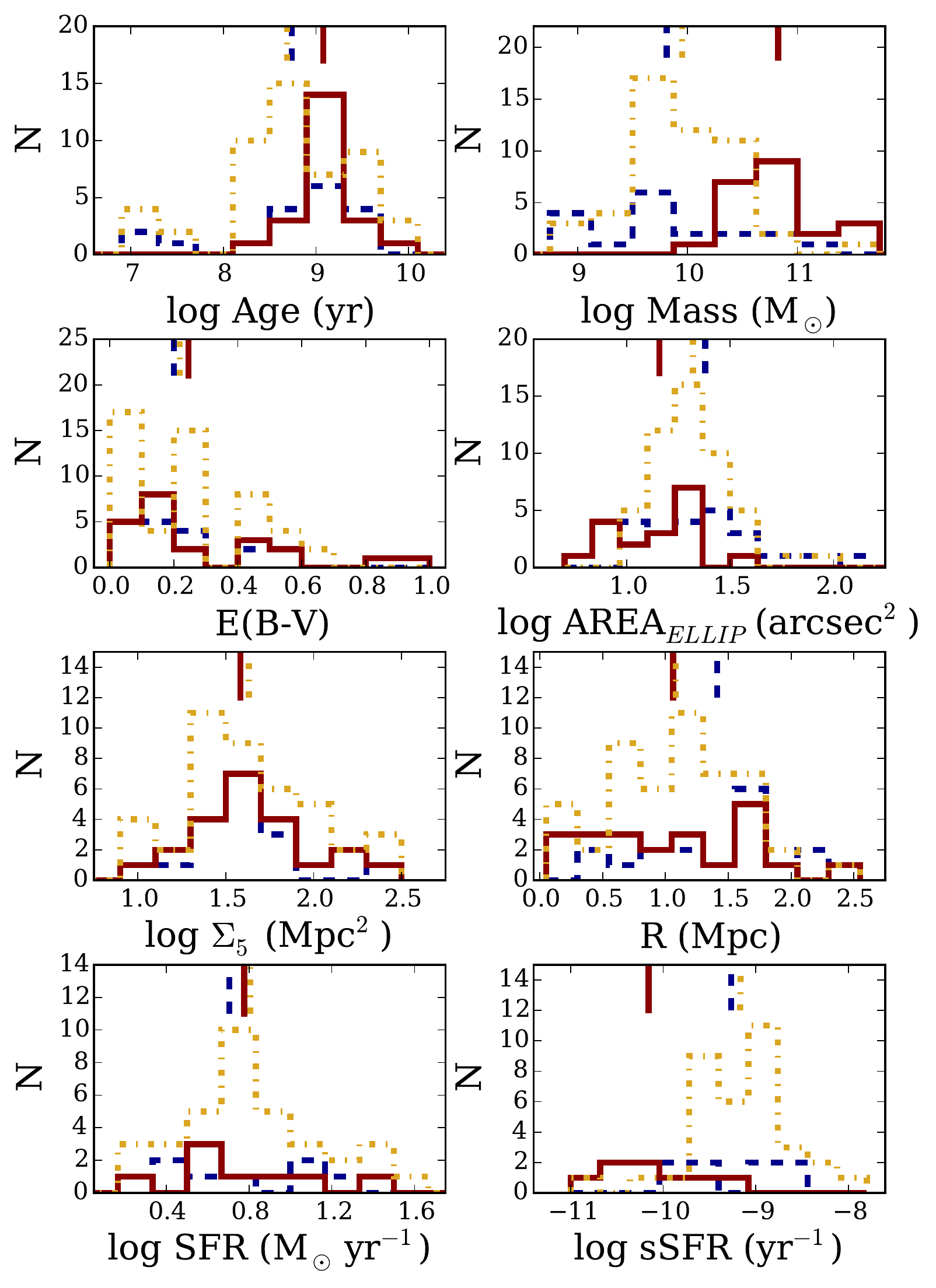}
\caption{Distributions of the physical properties for the three morphological samples: age, stellar mass, E(B-V), size, local density, cluster--centric radius, SFR, and sSFR. The LT sample is represented by the dark blue dashed line, the ET sample by the red solid line, and the blue ET sample by the yellow dot-dashed line. Top vertical lines indicate the mean values of each distribution.}
\label{fig:morphoPropHist}
\end{figure}

Regarding their physical characteristics we observe in Fig.\,\ref{fig:morphoPropHist} the distributions of several physical parameters for each of the three morphological subsamples. In this figure, going from left to right and from top to bottom we show the age, stellar mass, and extinction (derived through SED--fitting, as explained in Sect.4.2 of Paper I); the size (roughly estimated as the area covered by the ellipse used by SExtractor to extract the source); the local density $\Sigma_5$ and the cluster--centric radius as measurements of the cluster environment; and, the SFR and sSFR (calculated through \cite{Kennicutt1998} calibrations, using the total FIR luminosity when FIR data is available and the \oii\ luminosity for the remaining emitters), to gauge the star formation activity. As shown by the top vertical lines in every panel, indicating the average value of each subsample, only the stellar mass and the sSFR seem to show a different behaviour: ET galaxies are more massive and less efficient forming stars than the LT and blue ET samples. A Kolmogorov--Smirnov test confirmed this difference, since the statistics gives a probability of almost zero that the LT/``blue'' ET and the ET distributions are drawn from the same parent population. Nevertheless, for the three morphological types, the average value of these physical parameters in included within the ranges defined by the average value plus/minus its standard deviation of the other morphological types, suggesting that the difference is not so significant. Considering the peak of the distributions it seems that the blue ET population is shifted towards younger ages, however none statistical analysis confirm this fact. In relation with the environment, we find a difference in the cluster--centric distance distribution, where most of the LT galaxies are placed at distances higher than 1.5\,Mpc. 

\begin{figure}
\centering
\includegraphics[width=\hsize]{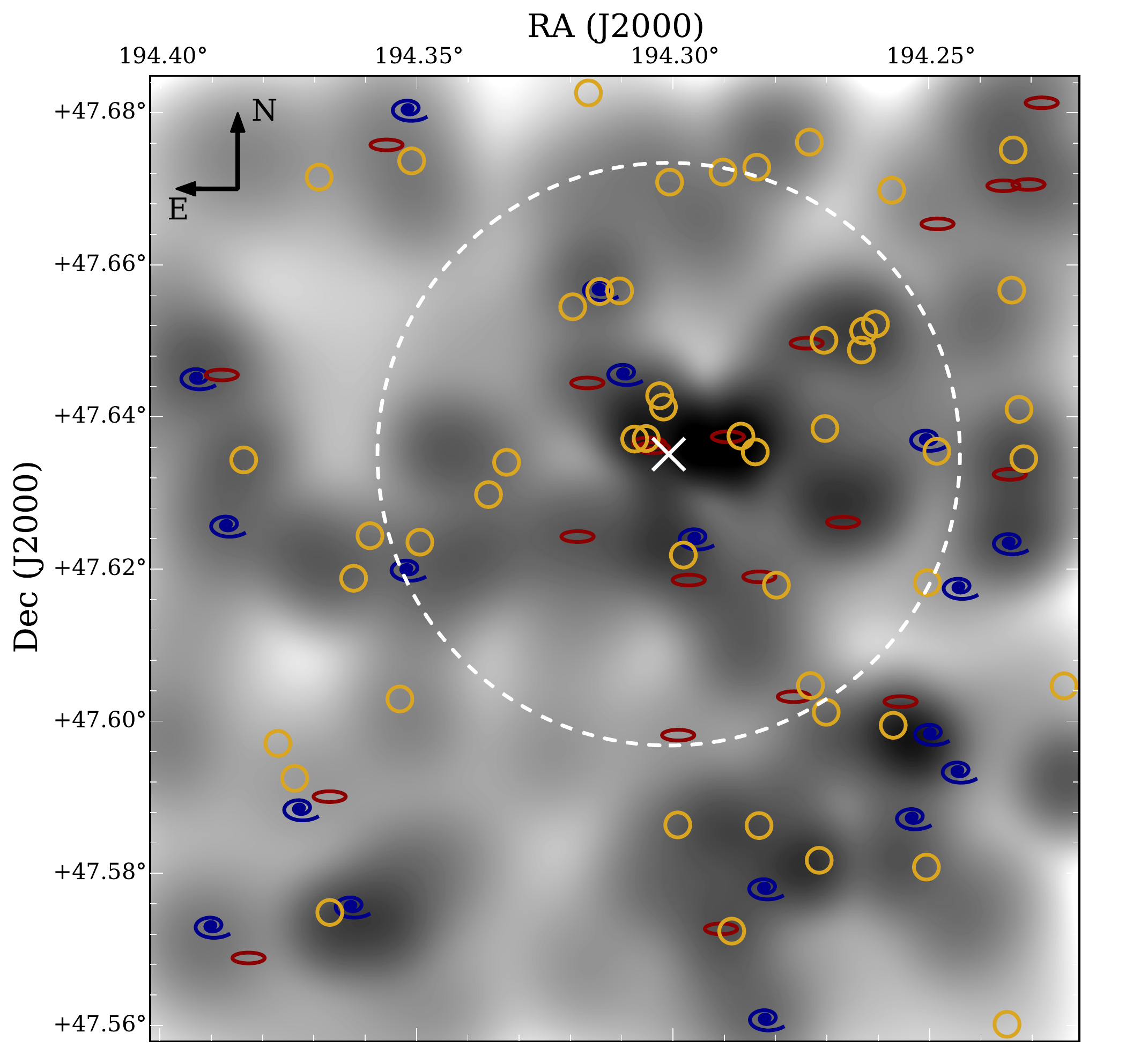}
\caption{Projected spatial distribution for the three morphological samples represented over the cluster density map. The white cross and dashed white circle are the cluster centre and virial radius, respectively. Symbols are the same as in Fig.\,\ref{fig:morphoColMag}.}
\label{fig:sptDist}
\end{figure}

Figure\,\ref{fig:sptDist} shows the cluster galaxy density map with the projected spatial distribution of the three morphological samples. Looking this spatial distribution we do not find any definitive trend. As observed through the cluster--centric distance distributions in Fig.\,\ref{fig:morphoPropHist}, more than 70\% of the LT sample is outside the virial radius, while about half of the ET populations are inside this radius. For the galaxy density we found that the three categories of galaxies show no preferred location, since they appear both in dense groups and as isolated galaxies. Qualitatively, LT galaxies are the only population avoiding the cluster core region and LT neighbours. On the contrary, the blue ET population do appear in high densities and groups that include any morphological type. Then, there is also no relevant difference in the spatial distribution of the blue ET galaxies respect to the ET sample. We have checked that no evidence appear if we estimate the local density with only the closest neighbour or if we include up to the tenth neighbouring galaxies.

Finally, looking the number of galaxies of each morphological type that show \oii\ or FIR emission, the percentage of ``blue'' ET galaxies with active star formation is significantly high: while star--forming galaxies in the ET and LT subsamples are less than a half of their population, the ``blue'' ET galaxies with ongoing star formation represent the 70\%. In Table\,\ref{tab:morphoEmi} we show the numbers and percentages of the \oii--detected and FIR--detected galaxies, and of the emitters sample, for the three morphological types. The fact that 70\% of the ``blue'' ET sample are star--forming galaxies supports the finding that the $g'$\,-\,$z'$ colour is so blue. In the following section we will further investigate the origin of this population.

		\begin{table}
	    \caption{Number of LT, ET and blue ET galaxies with emission (FIR-- and/or \oii--detected), indicating the percentage relative to each morphological type.}
	    \label{tab:morphoEmi}
		\begin{center}
		\begin{tabular}{l c c c}
		\hline\hline
		  & {ET} & {LT} & {blue ET} \\
		\hline
         \oii\ & 2 (9\%) & 7 (39\%) & 26 (52\%) \\
         FIR & 6 (27\%) & 3 (17\%) & 18 (36\%) \\
         \hline
         Emitters & 8 (36\%) & 8 (44\%) & 33 (70\%)   \\
        \hline
		\end{tabular}
        \begin{flushleft}
        \end{flushleft}
		\end{center}
		\end{table}

\section{The blue ET population}
\label{sec:blueET}
We have observed that, in general, ET, LT and ``blue'' ET samples within the cluster show similar average physical properties. Therefore, the existence of such a ``blue'' ET population could question its classification. In this section we will analyse possible explanations for observing such population. These explanations include that: 
\begin{itemize}
\item Morphological classification could be incorrect,
\item Redshift estimation could be incorrect,
\item Galaxies could be related with peculiar features,
\item Galaxies could host an AGN,
\item The blue colours could come from photoionising post--AGB stars,
\item They could be ET galaxies with current/recent star formation, and finally,
\item They could be luminous compact blue galaxies (LCBG) or late--type galaxies with enhanced, centrally concentrated star formation.
\end{itemize}

In the following subsections we discuss each of these possibilities and the analysis that we have performed.

\subsection{Unreliable morphological classification}
First, it is possible to question the performed morphological classification; however we have been very restrictive in its definition. Although some of the ``blue'' ETs might still be misclassified LT galaxies, most should be earlier types since they have very high probabilities. More than 45\% of them have probabilities of being ET above 0.8, and only 3 sources show probabilities below 0.7. 

In order to test the reliability of our morphological classification, and to prove that most of the ``blue'' ETs are not LT galaxies incorrectly classified, we have applied the same statistical methodology to a cluster at similar redshift and using comparable imaging data that has a trustful morphological classification. Finding such a cluster is not a trivial task. 
The selected cluster was MS1054-03 at z\,=\,0.83. For this cluster \citet{VanDokkum2000} performed a visual morphological classification based on WFPC2/HST images in the F606W and F814W bands, and \citet{Tran2005} carried out a spectroscopic campaign with LRIS/Keck confirming 121 cluster members within the WFPC2/HST mosaic. Later this cluster was observed with the ACS/HST using F606W, F775W, and F850LP bands, and \citet{Postman2005} performed a new visual morphological classification based on these images. 
There is a publicly available catalogue from \citet{Blakeslee2006} that includes the morphological type and spectroscopic redshift of 142 members of this cluster Regarding the available imaging data we explored two options: i) the F606W WFPC2/HST available science mosaic \citep{Forster2006}, and ii) the Rc SuprimeCam/Subaru raw images available under request at the SMOKA Science Archive. For the first option, it is necessary to first add background noise\footnote{we used the IRAF task \texttt{mknoise}, including the HST gain value and activating the \texttt{poisson} option for the noise.} and then degrade to our $r'$--band image resolution\footnote{we use the task \texttt{tgauss} within the IRAF dedicated package TFRED to equate the WFPC2 image to our pixel scale of 0.254$''$ and our seeing $\sim$\,0.9}. For the second dataset we followed a standard data reduction process using the SDFRED1 software \citep{Yagi2002,Ouchi2004}. 
In order to obtain comparable data we used only two individual exposures, which gave us a completeness magnitude at 50\% of Rc\,=\,24.5\,mag. For SuprimeCam the pixel scale is 0.2$''$/px, and the measured seeing was compatible with our value of 0.9$''$.  
Finally, we used these F606W and Rc processed images to execute the galSVM code, following the same methodology employed for the RXJ1257 galaxies (as detailed in Sect.\,\ref{subsec:morphoClas_method_real}). The statistical morphological classification was performed only for those galaxies with spectroscopic redshift (thus ensuring their cluster membership) with existing visual morphological classification from \citet{Postman2005}; furthermore we restricted the selection up to 25\,mag in all bands. These conditions resulted in a sample of 137 cluster galaxies, morphologically classified with a visual methodology and detected in both HST and Subaru images. In Table\,\ref{tab:ms1054} we show the results obtained for each image and magnitude bin, indicating for each type the number of galaxies for which the statistical and visual classification match (Good), the number of galaxies for which the statistical method fails by classifying the galaxy in other morphological type (Wrong) or by not being able to classify it (Uncl.). From the visual morphological classification we included as LT those galaxies classified between 1 and 10, and as ET those galaxies with types between 0 and -5. Note that in this table we have omitted the fainter bin (24.5\,$<$\,m\,$\leq$\,25), so the total number of objects shown for both HST and Subaru results is less than the full sample of 137 cluster galaxies. The numbers shown correspond to a standard probability threshold of $p_{ET}-\varepsilon(p_{ET})$\,$\geq$\,0.6 to be classified as ET and $p_{ET}+\varepsilon(p_{ET})$\,$\leq$\,0.4 to be classified as LT. From this table, it is evident that the results derived from the HST degraded image are better than those obtained with the Subaru image. This is not surprising since, although its completeness magnitude and resolution are comparable with our $r'$ image, the signal--to--noise is worse being around two times higher for the RXJ1257 $r'$ image. The main result that we want to test at this point is the possible fraction of wrongly classified LT galaxies: are most of the ``blue'' ET galaxies misclassified LTs and are we able to recover LTs? From the numbers in Table\,\ref{tab:ms1054} we obtain that the percentage of LT galaxies misclassified as ET goes from  8(3)\% for the better quality HST image to 15(6)\% for the poorer quality Subaru image when considering it with respect to the number of LT galaxies morphologically classified (when considering it with respect to the total number of LT galaxies). From these numbers, we can sustain that the existence of such ``blue'' ET population is not caused by an incorrect classification of a large part of the LT population.

		\begin{table}
	    \caption{Classification results for the MS1054-03 cluster obtained for both HST and Subaru images, per magnitude bin.}
	    \label{tab:ms1054}
		\begin{center}
		\begin{tabular}{L{1.2cm} C{1.cm} C{1.cm} C{1.cm}}
        \multicolumn{4}{c}{\textbf{WPFC2/HST F606W image}} \\
		\hline\hline
		 Morpho & Good & Wrong & Uncl. \\
		\hline
        \multicolumn{4}{c}{F606W\,$\leq$\,23} \\
        \hline
         LT\,=\,5 & 4 & 0 & 1 \\
         ET\,=\,9 & 2 & 1 & 6 \\
        \hline
        \multicolumn{4}{c}{23\,$<$\,F606W\,$\leq$\,23.5} \\
        \hline
         LT\,=\,10 & 2 & 0 & 8 \\
         ET\,=\,16 & 1 & 0 & 15 \\
        \hline
        \multicolumn{4}{c}{23.5\,$<$\,F606W\,$\leq$\,24} \\
        \hline
         LT\,=\,9 & 3 & 1 & 5 \\
         ET\,=\,27 & 10 & 0 & 17 \\
        \hline
        \multicolumn{4}{c}{24\,$<$\,F606W\,$\leq$\,24.5} \\
        \hline
         LT\,=\,8 & 3 & 0 & 5 \\
         ET\,=\,23 & 0 & 7 & 16 \\
        \hline\hline
        \vspace{0.05cm} \\
        \multicolumn{4}{c}{\textbf{SuprimeCam/Subaru Rc image}} \\
		\hline\hline
		 Morpho & Good & Wrong & Uncl. \\
		\hline
        \multicolumn{4}{c}{Rc\,$\leq$\,23} \\
        \hline
         LT\,=\,12 & 4 & 1 & 7 \\
         ET\,=\,29 & 7 & 10 & 12 \\
        \hline
        \multicolumn{4}{c}{23\,$<$\,Rc\,$\leq$\,23.5} \\
        \hline
         LT\,=\,9 & 6 & 0 & 3 \\
         ET\,=\,23 & 4 & 9 & 10 \\
        \hline
        \multicolumn{4}{c}{23.5\,$<$\,Rc\,$\leq$\,24} \\
        \hline
         LT\,=\,7 & 1 & 1 & 5 \\
         ET\,=\,20 & 6 & 2 & 12 \\
        \hline
        \multicolumn{4}{c}{24\,$<$\,Rc\,$\leq$\,24.5} \\
        \hline
         LT\,=\,5 & 0 & 0 & 5 \\
         ET\,=\,19 & 4 & 3 & 12 \\
        \hline\hline
		\end{tabular}
        \begin{flushleft}
        \end{flushleft}
		\end{center}
		\end{table}

\subsection{Unreliable cluster membership}
Furthermore it is possible to cast doubts on the derived cluster membership, i.e. that these sources are at different redshifts. Nevertheless, the estimation of each source redshift to build the cluster members catalogue has been done carefully: we have spectroscopic redshifts for 21 sources (Paper I), we obtain pseudo--spectroscopic redshifts for 87 \oii--emitters (118 without v$_{LoS}$ constrains, Paper III), and we estimated photometric redshifts within the cluster for 271 galaxies (Paper I) from which 64 were confirmed by the \oii--emission line. In addition, as Table\,\ref{tab:morphoClas} shows, more than half of the ``blue'' ET population are classified with the \oii--emission line and similar percentages ($\sim$70\%) of the photometric redshift sources are found in the three morphological types. We have compared the physical properties of the blue ET galaxies that were classified as a cluster member with their pseudo--spectra with those that were classified through photometric redshifts, i.e. blue ET galaxies \oii--detected and --undetected. The distributions of both samples are compatible supporting the soundness of the cluster membership classification.

\subsection{Galaxies with peculiar features}
Some of previous works suggested that many galaxies falling into clusters may evolve directly from peculiar, merging, and compact systems into ET galaxies, without passing through the phase of regular spirals. \cite{Nantais2013} observed this by studying the morphology with HST data of a cluster at redshift z\,=\,0.84, which is similar to ours. In this case, taking into account the poor resolution of our data, we could be missing the signs of peculiar structures, detecting therefore galaxies as earlier types, but their colours will be bluer in comparison with the standard elliptical and lenticular sources. To test this, and how well we are able to detect the peculiar structures with our data, we analysed the distribution of used morphological parameters on a sample of peculiar sources simulated to our conditions. We selected a sample of 300 local galaxies from \cite{Nair2010} catalogue, classified by authors as galaxies with unusual forms and with T-Type\,=\,-99. We simulated these galaxies to map the conditions of our data, as explained in Sect.\,\ref{sec:morphoClas_method}, and we measured the same morphological parameters. We repeated this process in 5 magnitude cuts, corresponding to those analysed in Sect.\,\ref{subsec:morphoClas_method_real}. Finally, we compared the distributions of obtained parameters with the same ones measured for out data, to check if they map the same space of values. We run the Kolmogorov-Smirnov test, obtaining in all cases that the two distributions are significantly different. Therefore, although some of the blue ET galaxies might enter in the category of peculiar or merging systems transforming to ETs directly, we discard the possibility that this is the case of a majority of them.   

		\begin{table}
	    \caption{LT, ET and blue ET galaxies that were classified as cluster member by its spectrum (GMOS), by SED--fitting (photometric bands), and/or by its pseudo--spectrum (TF/OSIRIS).}
	    \label{tab:morphoClas}
		\begin{center}
		\begin{tabular}{l c c c}
		\hline\hline
		 Method & {ET} & {LT} & {blue ET} \\
		\hline
         Spectrum  & 6 (23\%) & 2 (11\%) & 2 (4\%)  \\
         Pseudo--spectrum & 2 (9\%) & 7 (39\%) & 26 (52\%)   \\
         Photometry & 15 (68\%) & 13 (72\%) & 38 (76\%)   \\
        \hline
		\end{tabular}
        \begin{flushleft}
        \end{flushleft}
		\end{center}
		\end{table}

Taking into account the previous three points and the high number of galaxies classified as ``blue'' ET, we may consider that the bulk of them form a different population. As mentioned above, this population could host an AGN, could consist of ET galaxies with recent star formation activity, and/or could be related with the LCBG \citep{Crawford2014} population. 

\begin{figure}
\centering
\includegraphics[width=\hsize]{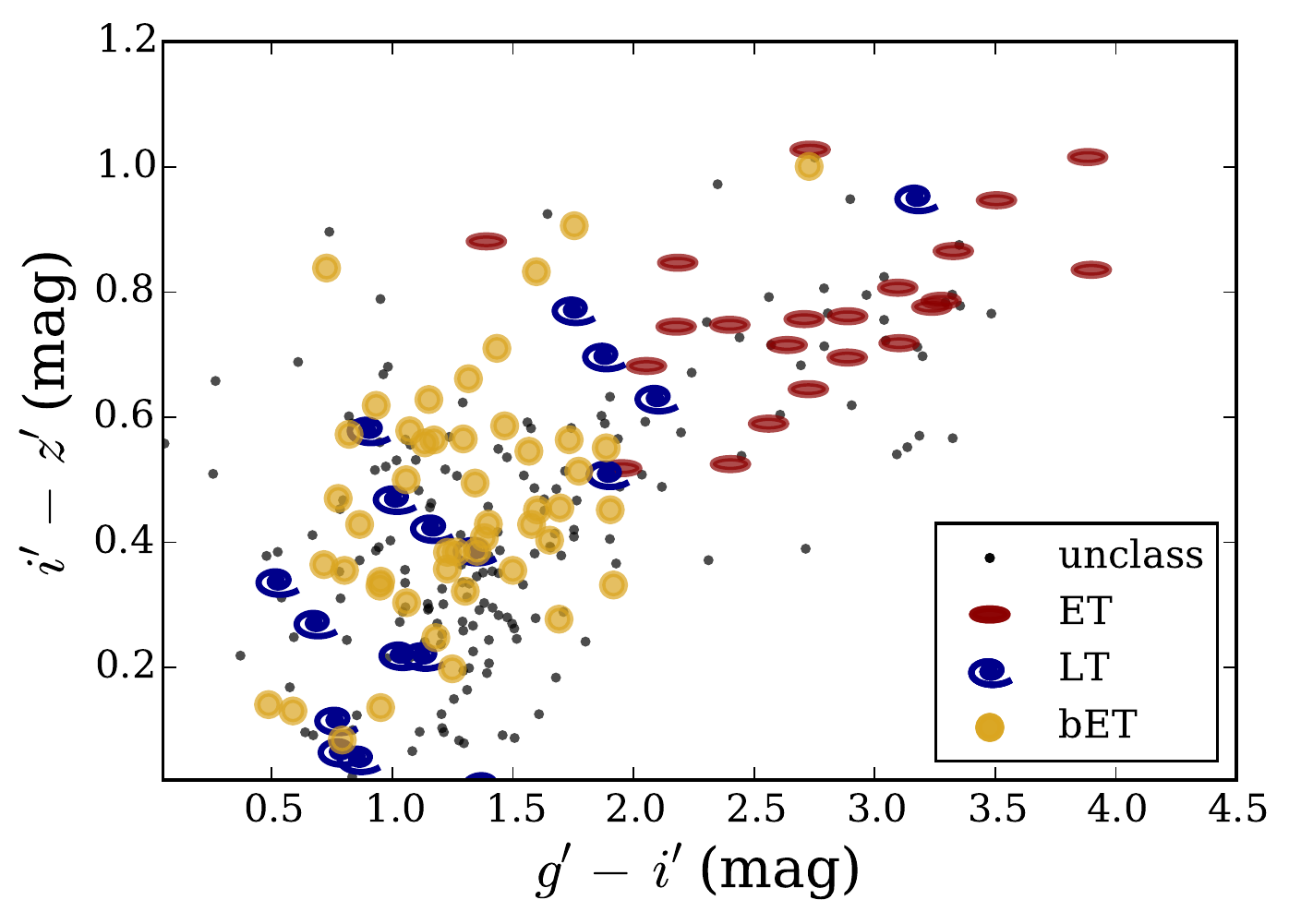}\\
\caption{colour--colour diagram of RXJ1257. Symbols are the same as in Fig.\,\ref{fig:morphoColMag}: blue spirals represent galaxies classified as LT, red ellipses as ET, yellow circles are ``blue'' ET galaxies, and small black dots are morphologically unclassified cluster members.}
\label{fig:morphoColgiz}
\end{figure}

Figure\,\ref{fig:morphoColgiz} shows the $g'-i'$ vs. $i'-z'$ colour--colour diagram of the RXJ1257 cluster members. Following \citet{Smail1998}, three types of galaxies could be analysed in this kind of diagram: 1) those with red colours, typical of passive elliptical galaxies, 2) those with blue colours typical of star-forming galaxies, and 3) those with red $i'-z'$ colour indicating a strong 4000\,\AA\ break and presence of older stellar populations, but with bluer $g'-i'$ colour then in the case of group 1). The population of $\sim$\,40\% of the blue ET galaxies falls above the threshold $i'-z'$\,=\,0.5\,mag, where we find the galaxies classified as standard ETs (with red colours), indicating that a large fraction of the sample shows the presence of evolved stellar populations, but with bluer $g'-i'$ colour than red ETs. This opens the possibility that these galaxies are, or recently have been, forming stars. In the ``transition'' region between 0.5 and 0.3\,mag in the $i'-z'$ colour, we found other $\sim$\,40\% of blue ETs, while only $<$\,20\% of blue ETs are located in the bluest region.

\subsection{AGN galaxy hosts}
One possibility that cannot be ruled out is that at least a fraction of the blue ET galaxies host an AGN; the nuclear contribution could make the integrated colours bluer and eventually place the host galaxies in the blue area of the CMD.  According to the estimations of \cite{Lemaux2010}, derived from \oii\  and \Ha\ measurements of two clusters at similar redshift (RXJ1821.6+6827at z\,$\approx$\,0.82 and Cl1604 at z\,$\approx$\,0.9), a substantial percentage of the ELG population, $\sim$\,68\% can be classified as AGNs and that nearly half of the sample have \oii\ to \Ha\ EW ratios higher than unity, the typical value observed for star-forming galaxies. At lower redshift (ZwCl 0024.0+1652 at z\,$=$\,0.395) our team \citep{SanchezPortal2015} derived, using measurements of the \Ha\ line width and \niiHa\ ratios, a more modest, but also substantial percentage  of AGNs, around 37\%.  At this stage of the survey of RXJ1257 we have only measurements of the \oii\ line and therefore we cannot derive the AGN fraction from diagnostics based on emission lines \citep[e.g.][]{Rola1997}. We have four-channel IRAC photometry for eleven blue ET galaxies and therefore we can apply the MIR AGN selection criterion proposed by \cite{Stern2005} based on the IRAC colours 3.6\,-\,4.5 and 5.8\,-\-8.0, to get a rough idea of the importance of the AGN contamination. Of course, given the skinny sample, the results should be taken with caution.  According to the authors, overall the criterion provides a robust separation of AGNs from SF galaxies, with 80\% completeness and less than 20\% contamination. Applied to our sample, a single galaxy (that also happens to be an \oii\ emitter) out of eleven falls within the AGN region of the colour--colour space. Therefore, a crude low-limit of 10\% of the population of blue ET galaxies could be guessed as hosting an AGN.

\subsection{Photoionisation by post--AGB stars}
One possibility that cannot be ruled out is that post--asymptotic giant branch (post--AGB) stars, with T$_{eff}$\,$\sim$\,10$^5$\,K, might be at least partially responsible of the blue colours. According to \cite{Binette94}, the photoionisation by post--AGB stars can account for the extended ionized gas often observed in elliptical galaxies. \cite{Ho2008} estimated that the nebular line emission in roughly one--third of the sources in their sample of nearby active galaxies can be powered by photoionisation from post--AGB stars. The percentage of LINERs and transition objects that could be powered by this mechanism is outstanding, 39\% and 33\% respectively \citep[see also][]{CidFernandes2011}.

\subsection{SF or LCB galaxies}
\begin{figure}
\centering
\includegraphics[width=\hsize]{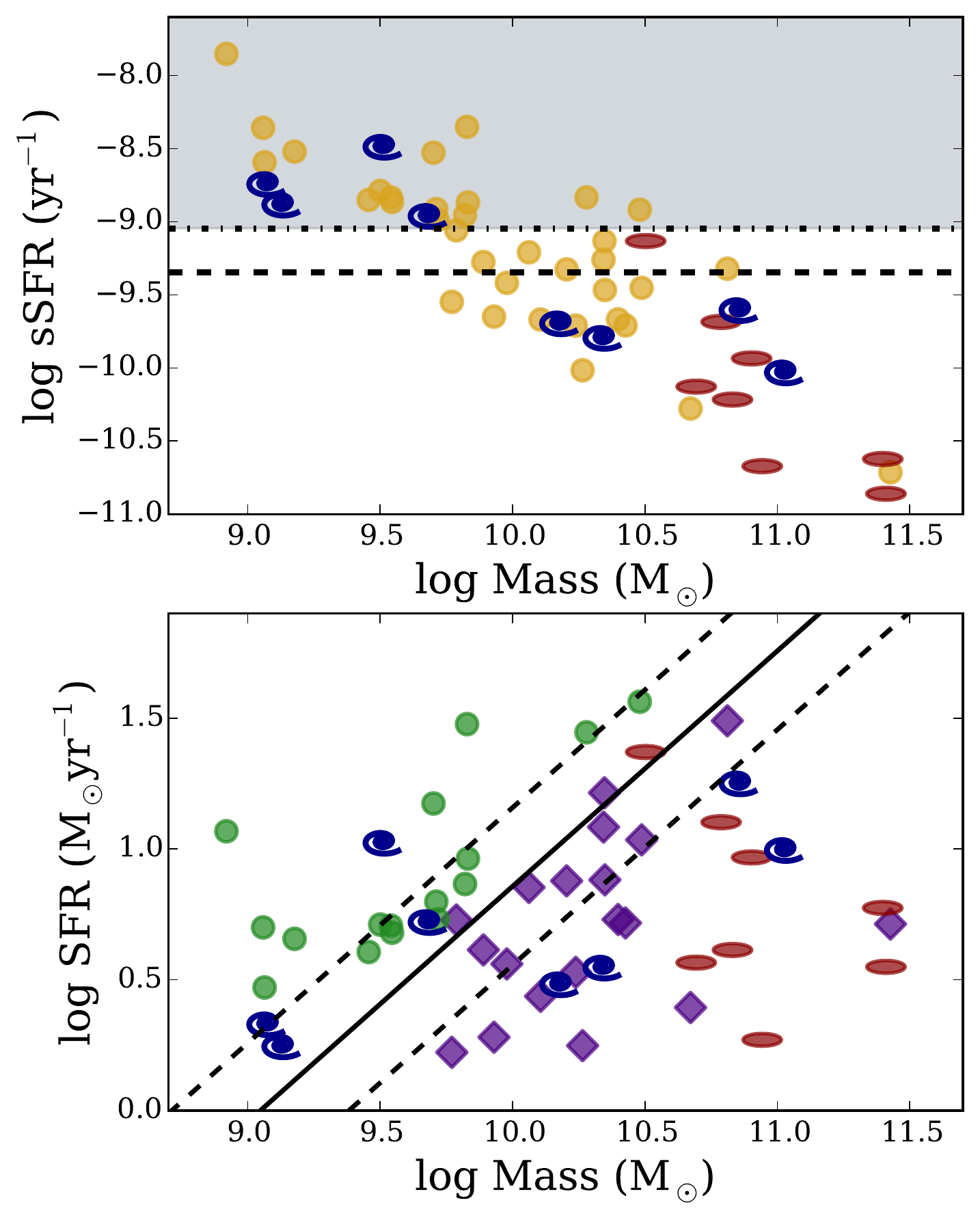}
\caption{sSFR (\textit{top}) and SFR (\textit{bottom}), estimated both with the FIR and \oii\ emission, as function of the stellar mass. As in previous figures, dark blue spirals are LT and red elongated ellipses are ET galaxies.  In the top panel, yellow circles are ``blue'' ETs (those galaxies that by their probability are classified as ET but their $g'$\,-\,$z'$ colour is below the 2\,mag limit). The horizontal dashed line corresponds to the relation of \citet{Elbaz2011} for the main--sequence galaxies at the redshift of the cluster, and the dash--dotted line indicates the lower limit for the starburst region (gray shaded area). This line is the threshold used to split the sample into LCBGs (in the starbust area) and ETs with recent star formation (in the main--sequence). In the bottom panel, green circles are LCBGs and violet diamonds ET galaxies with  star formation, as classified in the top panel. The solid line indicates the SFR--M$_{\star}$ correlation from \citet{Elbaz2007}, and the dashed lines show the 68\% dispersion.}
\label{fig:morphoSFRMass}
\end{figure}

In order to analyse the last two options, we split the blue ET sample into what we are going to consider ``ETGs with recent SF'' and ``LCBGs'', but bearing in mind that this separation is purely speculative. We used the known sSFR--stellar mass correlation and the relation derived by \cite{Elbaz2011} for the main sequence galaxies to perform the sample division: we classified as ``LCBGs'' those galaxies placed in the starburst region, as shown in the upper panel of Fig.\,\ref{fig:morphoSFRMass} by the shaded area; and, as ``ETGs with recent SF'' those galaxies that show normal star--formation activity, being around and below the main--sequence relation, plotted with a dashed line in the upper panel of Fig.\,\ref{fig:morphoSFRMass}. This division would be equivalent if we used the SFR--stellar mass correlation from \cite{Elbaz2007}, as shown by the solid line in lower panel of Fig.\,\ref{fig:morphoSFRMass}. Since not every blue ET galaxy has detectable star formation activity, a third sample with the rest of ``quiescent'' blue ET galaxies is defined.

\begin{figure}
\centering
\includegraphics[width=\hsize]{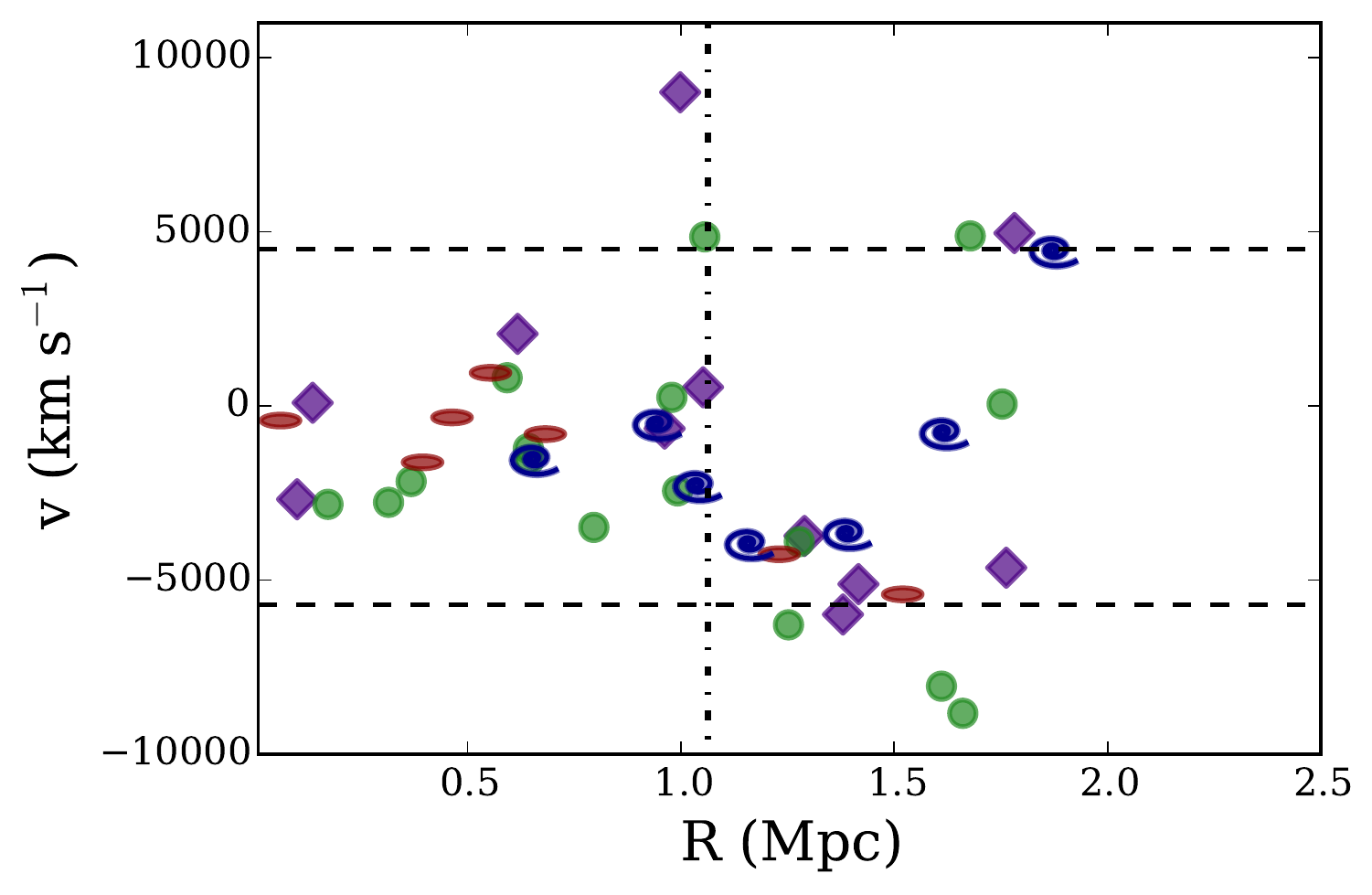}
\caption{Radius vs radial velocity for the morphological samples: ET (red ellipses), LT (blue spirals), ET with SF (violet diamonds), and LCBG (green circles). For the rest of the blue ET sample there is no dynamical information. The vertical dash--dotted line indicates the virial radius position, and horizontal dashed lines mark off the radial velocity range full covered within the OSIRIS field of view.}
\label{fig:vradR}
\end{figure}

\begin{figure}
\centering
\includegraphics[width=0.8\hsize]{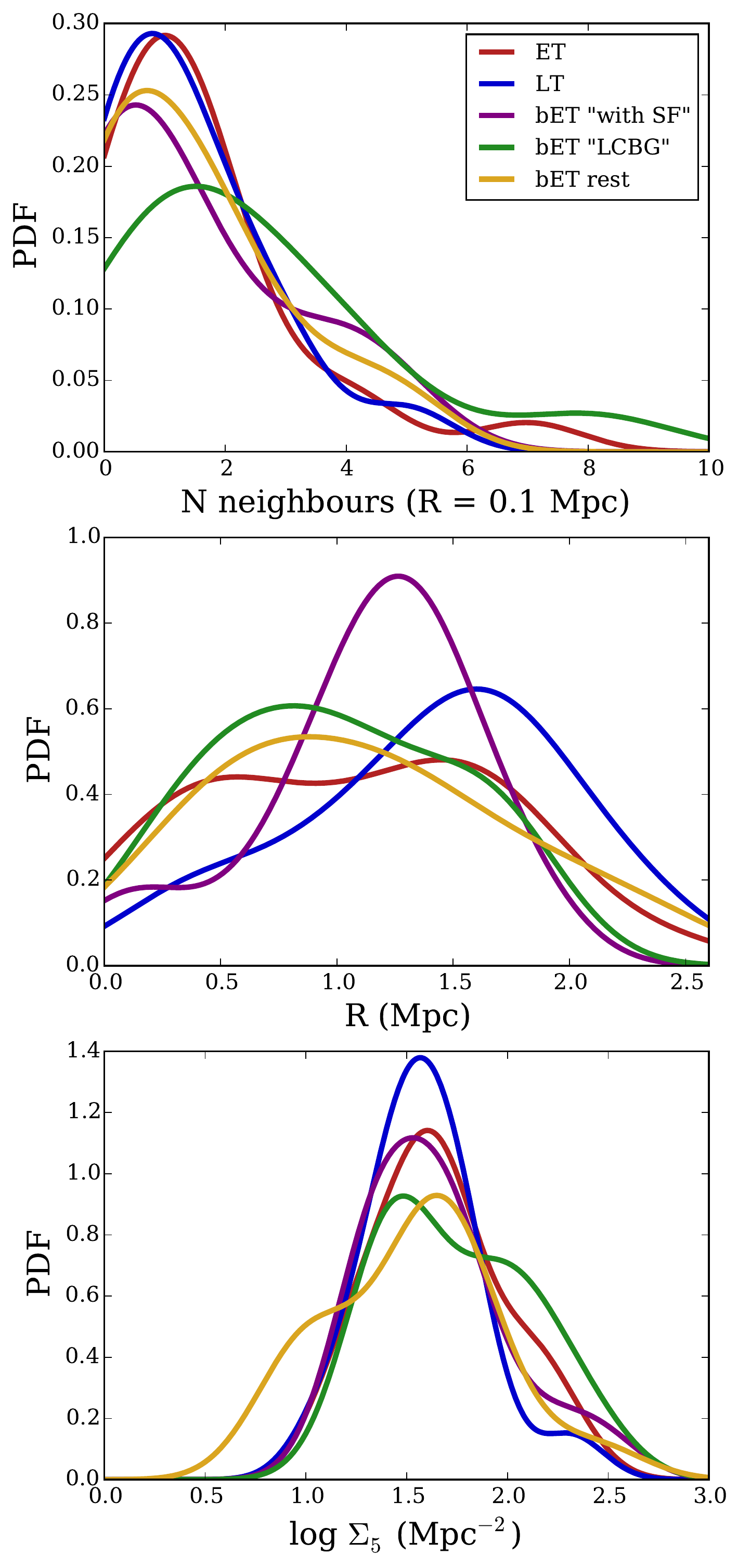}
\caption{Probability density function of the number of neighbours within a projected distance of 0.1\,Mpc (\textit{top}), the distance to the cluster core (\textit{centre}), and the local galaxy density (\textit{bottom}) for all the morphological types: ET (red), LT (dark blue), blue ET ``with recent SF'' (violet), blue ET ``LCBGs'' (green), and the rest of blue ETs (yellow). These curves are estimated by applying a Gaussian kernel density estimator to smooth the distribution.}
\label{fig:morphoNeigh}
\end{figure}

\citet{Crawford2014} analysed the spatial and kinematic distributions of the galaxy population of a sample of five clusters at intermediate redshift 0.5\,$<$\,z\,$<$\,0.9. The authors defined, based only on their colour, luminosity, and surface brightness, four galaxy samples: red sequence, blue cloud, green valley, and LCBGs. The evidences they found to support that the LCBG population defined was a different one are that LCBGs show a higher velocity dispersion, avoid the cluster core, and are more likely to occur in pairs or small groups, having their velocity similar to their near neighbours. To compare our blue ET population with these LCBGs we first analysed, in Fig.\,\ref{fig:vradR}, the kinematics for our four morphological samples, i.e. LTs, ETs, blue ETs ``with recent SF'', and ``LCBGs''. Figure\,\ref{fig:vradR} shows the radial velocity as a function of the cluster--centric radius, with the horizontal dashed lines indicating the velocity field full covered in the OSIRIS field of view (Paper III). Visually, there is no difference between the samples, however, we obtained that the average velocity value for the ``LCBG'' sample is significantly higher than the average values for the rest ($\sim$-2500\,km\,s$^{-1}$ for ``LCBGs'' versus $\sim$-1100, -880, and -560\,km\,s$^{-1}$ for LTs, ETs, and blue ETs ``with recent SF'', respectively). The standard deviation values are compatible fot the four samples, showing all a high dispersion with values between 2600 and 4400\,km\,s$^{-1}$. Analyzing Fig.\,\ref{fig:morphoNeigh}, we investigate the environment of each population. Figure \ref{fig:morphoNeigh} show the probability density function (i.e. the original distribution smoothed with a Gaussian kernel density estimator) of the number of nearest neighbours, distance to the cluster core, and local density for the five morphological populations (LTs, ETs, blue ETs ``with recent SF'', ``LCBGs'', and the rest of the blue ETs, those we could consider as ``quiescent''). Although no dramatic differences between these populations are observed, there are some notable facts: (\textit{i}) the curves of the ``LCBG'' sample have always two peaks, which could indicate that is formed of two different populations; (\textit{ii}) regarding the number of neighbours within 0.1\,Mpc of distance the ET and LT populations are the most isolated galaxies, while the ``LCBGs'' seems to be the most grouped, as we could expect from the \cite{Crawford2014} study; (\textit{iii}) blue ET ``with recent SF'' prefer the cluster distance just above the virial radius; and, (\textit{iv}) there is no difference in local galaxy density except for the second peak of the ``LCBGs'' towards high densities and a large fraction of the ``quiescent'' blue ETs that appears at the lowest densities. Finally, there is a possibility that the true nature of blue ETs are LT galaxies, but with enhanced central star formation, appearing therefore as more compact and concentrated than normal LTs. They could be also responsible for the second peak of the ``LCBGs'', or could be classified into the ``ETs with recent SF'' group.

Considering that the number of galaxies with unreliable morphological classification or photometric redshift measurements is negligible (as explained above), the analysis that we carried out suggest that the population of galaxies classified as ``blue'' ETs is formed by ``true'' early-types, being compact and concentrated, although we are not able to describe their detailed structure and kinematics. This sample of ``blue'' ET galaxies probably includes a mixture of several populations: post--interacting systems with a presence of peculiar structures, AGN hosts, galaxies with current or recent star formation, and/or LCBG. At this point, new data are necessary to confirm the presence of each of these populations. However, being a young cluster and still in formation, RXJ1257+4738 opens the possibility for the existence of each of them, and presents a very good candidate for performing a more detailed morphological study in future.

\section{Morphology--density relation}
\label{sec:tsigmarelation}
In this section, we focus our attention in the relation between the morphology of the classified cluster galaxies and their environment. Although the sample is incomplete, we have seen in Fig.\,\ref{fig:morphoBias} that there is no  difference between the whole cluster and the morphologically classified samples in what concerns to the environment, i.e. the distributions of the local galaxy density and the distance to the cluster centre are comparable for the classified and unclassified populations. 

As observed in Fig.\,\ref{fig:morphoSFRMass}, where the relation between the star formation activity and the stellar mass was analysed, the ET sample is mostly located below the main--sequence, being only one galaxy in the sSFR--M$_\star$ diagram in the region of star--forming galaxies. We also found that half of the galaxies of the other two types are clearly placed in the star--forming region. Nevertheless, there is no clear relation between the morphological type and the average SFR, and the sSFR--M$_\star$ relation shows the stellar mass difference between massive ET and less massive LT and ``blue'' ET galaxies, indicating that stellar mass plays a more major role. Introducing the environment, calculated as the local galaxy density $\Sigma_5$, we analyze their dependency with those properties, the star formation activity and the stellar mass, for the different morphologies. We have observed that there is no tendency with environment, since the morphological types span the full local density range, and only the difference in stellar mass is seen (to show the plots is not worth since no trend become visible).

\begin{figure}
\centering
\includegraphics[width=\hsize]{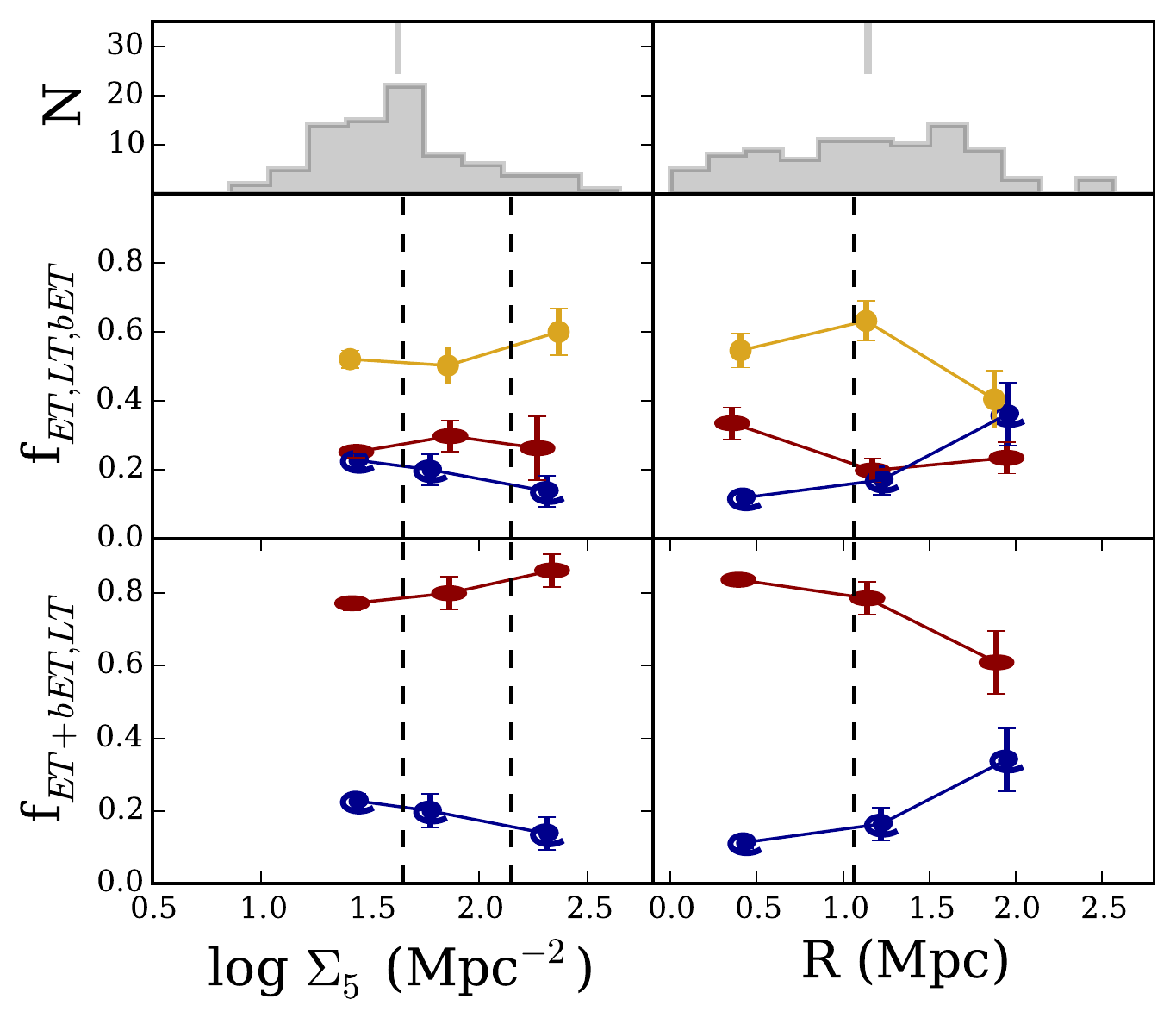}
\caption{Morphological types - LT (blue spirals), ET (red ellipses) and ET galaxies with blue $g'$\,-\,$z'$ colours (yellow circles) - as a function of the local galaxy density (\textit{left}) and the cluster--centric radius (\textit{right}), represented as the fraction of the morphological types in three densities/radius. Top panels represent the local density and cluster--centric distance distributions for the full classified sample. The vertical lines indicates the mean value of each distribution. The dashed vertical lines are the original local density limit between environments defined by KO08 (\textit{left}) and the virial radius estimated by \citet{Ulmer2009} (\textit{right}). Errorbars are estimated as the standard deviation of the fraction distribution built by bootstrap.}
\label{fig:morphoFrac}
\end{figure}

The direct relation between the cluster environment and each morphological type is approached now by analyzing the fraction of each morphological type as a function of the local galaxy density and the distance to the cluster centre. Environmental processes affect the morphological characteristics of galaxies, being  the morphology--density relation found by \citet{Dressler1980} a clear probe. In Fig.\,\ref{fig:morphoFrac} we show the fractions of each morphological type as a function of the local density (\textit{left}) and the cluster--centric radius (\textit{right}), considering galSVM probabilities and optical colours (\textit{top}) and only the galSVM classification (\textit{bottom}). The fraction values and their errors in each region were estimated performing a bootstrap analysis by randomly varying the $\Sigma_5$ and R limit values (see Paper I for more details). For the LT sample the trend is clear: they are scarce in high density regions and in the cluster centre, and their proportion grows, as we move towards less dense environments and larger distances from the cluster centre. In every density region studied, the cluster environment is evident since the number of LT galaxies is always below the other morphological types. The fraction of LT galaxies on the total of the morphologically classified galaxies barely reaches the 20\%, a result in contrast with, for example, \cite{Desai2007}. These authors found, in an intermediate redshift range of 0.5\,$<$\,z\,$<$\,0.8, that typical fractions for the LT sample are approximately the 55\%. Our work is difficult to compare with such kind of works because of the differences and sample incompleteness, nevertheless, as come into view in Fig.\,\ref{fig:morphoColMag}, we guess that the fraction of LT galaxies we obtained is highly biased since most of the unclassified cluster galaxies are placed in the fainter part of the blue cloud.

Regarding the ET samples, when we take them into account as a whole we find that they dominate the cluster core and the densest environments, and their proportion decrease opposite to the LT sample. However this evolutionary trend is really mild, specially considering the errorbars. When we distinguish by the optical colour between ET and ``blue'' ET galaxies that trend is not evident (we have to bear in mind the error bars). On the one hand, the different shape of the ET and ``blue'' ET fractions as a function of the local density and the cluster--centric radius reflects that the effect of the local environment and the cluster potential could not be the same.  \cite{Fasano2015temp} have recently found in a local cluster sample that, in fact, the morphology--density relation is only evident in the very inner region or in very regular clusters, while the morphology--radius relation remains unchanged for the whole sample, being only reduced for the really clumpy clusters. As discussed in Paper I, and as can be seen on the density map of Fig.\,\ref{fig:sptDist}, the RXJ1257 cluster is still under the process of collapsing and, therefore, presents a clumpy structure, which would explain the lack of a strong morphology--density relation. The poor number statistics make it impossible for us to check the morphology--density relation in the innermost part of the cluster. On the other hand, for the ET sample the trend of increasing its number towards high density regions and at the cluster centre is possible within the errors, and for the ``blue'' ET their fraction seems to grow at distances equivalent to the virial radius but also in the densest environments suggesting that this population may mainly appear as groups entering the cluster potential. Nevertheless, the fraction of ``blue'' ET galaxies is the highest in all environments at all distances, becoming the dominant population in this cluster at redshift 0.866.

\section{Summary and conclusions}
\label{sec:conclusions}
As part of a comprehensive study of the galaxy cluster RXJ12657+4738, we have performed in the present work a morphological analysis of its cluster galaxies. Taking into account both, the high redshift of the cluster and the low resolution of our optical broad--band data, we decided to test the non--parametric methods in order to classify our galaxies morphologically. We used the galSVM code, which was tested successfully in previous works in both cases, and simulated a set of 4000 local, visually classified galaxies to the conditions of RXJ1257 in order to classify the galaxies from our cluster. Our classification is statistical, providing for each galaxy the probability that it belongs to early-- (E/S0) or late--type (spirals and irregulars). In absence of better data and/or other morphological classifications, to test our classification and define the probability thresholds that ensure a reliable classification, we analysed standard morphological diagnostic diagrams and physical properties of classified sources. Being strict with the probability limits, minimizing the possibility of wrong classification, we classified 90 galaxies in total, $\sim$\,30\% of our whole sample, out of which 72 are early--types and 18 are late--types.  

In the observed CMD, a large fraction of ET galaxies unexpectedly falls in the blue cloud. We tested on another cluster with similar properties and with HST/ACS visual morphological classification available, that we are able to recover LT galaxies. Therefore we can trust that ETs with bluer colours are really early-types rather than misclassified sources. We analysed the physical properties of this blue ET galaxies separately. \\
The main results of our analysis are:
\begin{itemize}
\item The comparison, among the three morphological types, of the morphological parameters, the SED--derived properties (e.g. stellar mass), and other physical properties (as those related with the environment or the star formation activity, e.g. $\Sigma_5$ or SFR) shows no clear differences. Only the stellar mass and sSFR mean values suggest that the ET population seems to be more massive and less efficient forming stars than the LT and ``blue'' ET samples.
\item Such a blue optical colour is supported by the high percentage of ``blue'' ET galaxies forming stars, 70\% of the sample.
\item Neither a mistaken morphological classification nor an incorrect redshift estimation could explain the high number of galaxies classified as ``blue'' ET. These objects have ET morphologies in the sense that they are compact and concentrated, although their detailed structure and kinematics remain as open question.
\item The lack of a significant difference on the properties analysed (as the spatial distribution, environment, radial velocity, or star formation activity) suggests that blue ETs might consist of a mixture of different populations. The possibilities that we analysed include: (i) galaxies with a peculiar morphology transforming directly to ETs without passing the phase of regular spirals, (ii) ETs hosting an AGN, where nuclear contribution could make the integrated colours bluer, (iii) photoionisation by post--AGB stars, (iv) ETs with recent or remaining star formation, (v) LCBGs, and (vi) LTs with star formation activity highly concentrated in their central parts. The violent environment of this cluster under the process of formation could favour the presence of these named populations.
\item The analysis of the morphology--density and the morphology--radius relations becomes uncertain since, not only the number statistics of morphologically classified galaxies are poor, but also most unclassified galaxies fall into the fainter part of the blue cloud (see Fig.\,\ref{fig:morphoColMag}), which suggest that most of them would be LT galaxies. Despite this, when only ET/LT classification is considered we observed a mild morphology--density and morphology--radius relation. Nevertheless, when the separation by optical colour is taken into account these relations vanish. 
\end{itemize}

The significant presence of the blue early--type galaxies and lack of evident morphology--density and morphology--radius relations could be explained by the fact that RXJ1257+4738 is not a 'standard' cluster, but a young one, being still in the process of formation and showing a clumpy structure. The work we have carried out here leaves many open issues (as the origin of the ``blue'' ET population), but pointed our the interesting case that the RXJ1257+4738 cluster represents. There are no many similar examples in the literature to make constructive comparisons. However, we would like to stress once again, that all results presented in this paper in relation to morphology only reflect the situation of 30\% of our cluster members. Therefore, higher--resolution data are essential to obtain better and more complete classification and to shed some light on the unanswered questions.


\begin{table*}
\begin{center}
\caption{\label{tab:morphodata} Properties of the morphologically classified cluster galaxies (first seven rows). } 
\begin{tabular}{l c c c c c c c c c c c c}
\hline \hline
\multirow{2}{*}{ID} & RA J2000 & Dec J2000 & r$'$ & \multirow{2}{*}{Emission} &  \multirow{2}{*}{A}  &  \multirow{2}{*}{C}  &  \multirow{2}{*}{GINI} & \multirow{2}{*}{S} &  \multirow{2}{*}{M$_{20}$} &  \multirow{2}{*}{C$_{BC}$}  &\multirow{2}{*}{Probability} & \multirow{2}{*}{Morpho$^{a}$} \\
 & (12:mm:ss.ss) & (+47:mm:ss.s) & (mag) & & & & & & & & &  \\
\hline

452 & 56:53.71 & 36:16.9 & 23.17$\pm$0.05  & [O\textsc{ii}] & -3.72 & 0.41 & 0.60 & -1.14 & -0.98 & 6.65 & 0.96$\pm$0.12 & bET\\
      613 & 56:55.56 & 38:04.6 & 22.60$\pm$0.02  & [O\textsc{ii}]+FIR & 0.05 & 0.35 & 0.68 & - & -1.40 & 4.99 & 0.76$\pm$0.02 & bET\\
      685 & 56:56.09 & 39:24.3 & 23.98$\pm$0.07  & [O\textsc{ii}] & 0.06 & 0.31 & 0.57 & - & -1.43 & 5.49 & 0.87$\pm$0.08 & bET\\
      708 & 56:56.02 & 40:30.6 & 23.13$\pm$0.04  & [O\textsc{ii}] & 0.01 & 0.49 & 0.64 & 0.25 & -1.93 & 6.36 & 0.97$\pm$0.10 & bET\\
      892 & 56:58.59 & 37:04.0 & 24.20$\pm$0.07  & [O\textsc{ii}] & -0.01 & 0.26 & 0.67 & - & -1.03 & 3.81 & 0.20$\pm$0.05 & LT\\
      964 & 56:59.63 & 38:07.9 & 23.81$\pm$0.05  & [O\textsc{ii}] & 0.00 & 0.28 & 0.55 & - & -1.18 & 4.83 & 0.76$\pm$0.02 & bET\\
      974 & 57:00.09 & 37:06.1 & 22.55$\pm$0.02  & [O\textsc{ii}]+FIR & 0.13 & 0.33 & 0.65 & - & -1.13 & 4.99 & 0.72$\pm$0.03 & bET\\
      \hline
  		\end{tabular}		
        \begin{flushleft}
        \begin{footnotesize}
        \hspace{0.5cm}$^{a}$ LT: late--type, ET: early--type, and bET: blue early--type. 
        \end{footnotesize}
        \end{flushleft}
		\end{center}
\end{table*}

    \begin{acknowledgements}
We thank I. Smail for his helpful comments. We thank the anonymous referee for her/his close reading of the manuscript and her/his valuable comments. IPC acknowledges support from the Faculty of the European Space Astronomy Centre (ESAC). MP acknowledges financial support from JAE-Doc program of the Spanish National Research Council (CSIC), co-funded by the European Social Fund. This research has been supported by the Spanish Ministry of Economy and Competitiveness (MINECO) under grants AYA2014-58861-C3-1-P, AYA2014-58861-C3-3-P, AYA2011-29517-C03-01, AYA2010-15169 and AYA2013-42227-P, and by the Junta de Andaluc\'ia under grant TIC 114. Based on observations made with the Gran Telescopio Canarias (GTC), installed in the Spanish Observatorio del Roque de los Muchachos of the Instituto de Astrof\'isica de Canarias, in the island of La Palma, as part of the large ESO/GTC program ID.\,186.A-2012. This research made use of Astropy, a community--developed core Python package for Astronomy (Astropy Collaboration, 2013).
   
    \end{acknowledgements}

\bibliographystyle{aa} 
\bibliography{referencias} 

\end{document}